\documentclass[twoside,12pt]{article}
\usepackage{epsfig}

\def\Journal#1#2#3#4{{#1} {#2} (#4) #3 }
\def\AJ{\em Astron. J.}
\def\APJ{\em Astrophys. J.}

\def\APJS{\em Astrophys. J. Suppl. Ser.}
\def\AAP{\em Astron. Astrophys.}
\def\AAPS{\em Astron. Astrophys. Suppl. Ser.}
\def\MNRAS{\em Mon. Not. R. Astron. Soc.}
\def\NPA{{\em Nucl. Phys.} A}

\def\PLB{{\em Phys. Lett.} B}
\def\PRL{\em Phys. Rev. Lett.}
\def\PREV{\em Phys. Rev.}
\def\PREP{\em Phys. Rep.}

\def\PRD{{\em Phys. Rev.} D}
\def\PRC{{\em Phys. Rev.} C}

\def\RMP{{\em Rev. Mod. Phys.}}
\def\NAT{{\em Nature}}
\def\gtrsim{\:\lower 0.4ex\hbox{$\stackrel{\scriptstyle >}
{\scriptstyle\sim}$}\:}
\def\lesssim{\:\lower 0.4ex\hbox{$\stackrel{\scriptstyle <}
{\scriptstyle\sim}$}\:}
\def\sig{\:\lower 0.6ex\hbox{$\stackrel{\textstyle >}{\sim}$}\:}
\def\sil{\:\lower 0.6ex\hbox{$\stackrel{\textstyle <}{\sim}$}\:}

\def\S{Sec.\ }
\newcommand{\be}{\begin{equation}}
\newcommand{\ee}{\end{equation}}
\newcommand{\bea}{\begin{eqnarray}}
\newcommand{\eea}{\end{eqnarray}}

\begin{document}

\title{The Origin of the Heavy Elements: Recent Progress in\\
the Understanding of the $r$-Process}
\author{Yong-Zhong Qian\\
\\
School of Physics and Astronomy, University of Minnesota,\\
Minneapolis, MN 55455, USA}
\maketitle
\begin{abstract} 
There has been significant progress in the understanding
of the $r$-process over the last ten years. The conditions required for 
this process have been examined in terms of the parameters for adiabatic 
expansion from high temperature and density. There have been many 
developments regarding core-collapse supernova and neutron star merger 
models of the $r$-process. Meteoritic data and observations of metal-poor 
stars have demonstrated the diversity of $r$-process sources. Stellar
observations have also found some regularity in $r$-process abundance 
patterns and large dispersions in $r$-process abundances at low 
metallicities. This review summarizes the recent results from parametric 
studies, astrophysical models, and observational studies of the 
$r$-process. The interplay between nuclear physics and astrophysics is 
emphasized. Some suggestions for future theoretical, experimental, 
and observational studies of the $r$-process are given.
\end{abstract}
\vfill
\eject

\tableofcontents

\section{Introduction}
The basic framework for understanding the origin of the elements was
summarized in the classic works by Burbidge et al. \cite{b2fh} and 
Cameron \cite{al57} in 1957 (see \cite{wa97} for a more up-to-date 
review). The crucial observational basis for 
developing this framework was the abundance distribution of stable and 
long-lived nuclei in the solar system. The slow ($s$) and the rapid 
($r$) neutron-capture processes were proposed as the dominant 
mechanisms for producing the elements heavier than Fe. The nuclear 
systematics of these two processes provides a simple and beautiful
explanation for the peaks at mass numbers $A=80$, 88, 130, 138, 195, and 
208, respectively, in the solar abundance distribution (see Fig. 1).
The work by Seeger et al. \cite{se65} in 1965 gave a comprehensive
discussion of the $s$-process and the $r$-process based on the
underlying nuclear physics. More recent reviews of the $s$-process
can be found in \cite{ka89}--\cite{bu99}. This review focuses on the
progress in the understanding of the $r$-process since the last
major review by Cowan et al. \cite{co91} in 1991. After a basic
introduction, the astrophysical conditions for a successful 
$r$-process are discussed in \S2. Recent developments regarding 
core-collapse supernova and neutron star merger models of the 
$r$-process are reviewed in \S3. Data from studies of meteorites and 
metal-poor stars and their implications for the $r$-process are 
discussed in \S4. Some suggestions for future theoretical, experimental, 
and observational studies of the $r$-process are given in \S5.
For those readers who are interested in some particular aspects of 
the rich physics and astrophysics associated with the $r$-process, 
a good place to start is \S5, where a brief summary is also provided. 
Together with the table of contents, this summary may serve as a guide 
in finding the topics of special interest that are covered here.

\begin{figure}[tb]
\begin{minipage}{2 cm}
\makebox[1cm]{}
\end{minipage} 
\begin{minipage}{8 cm}
\epsfig{file=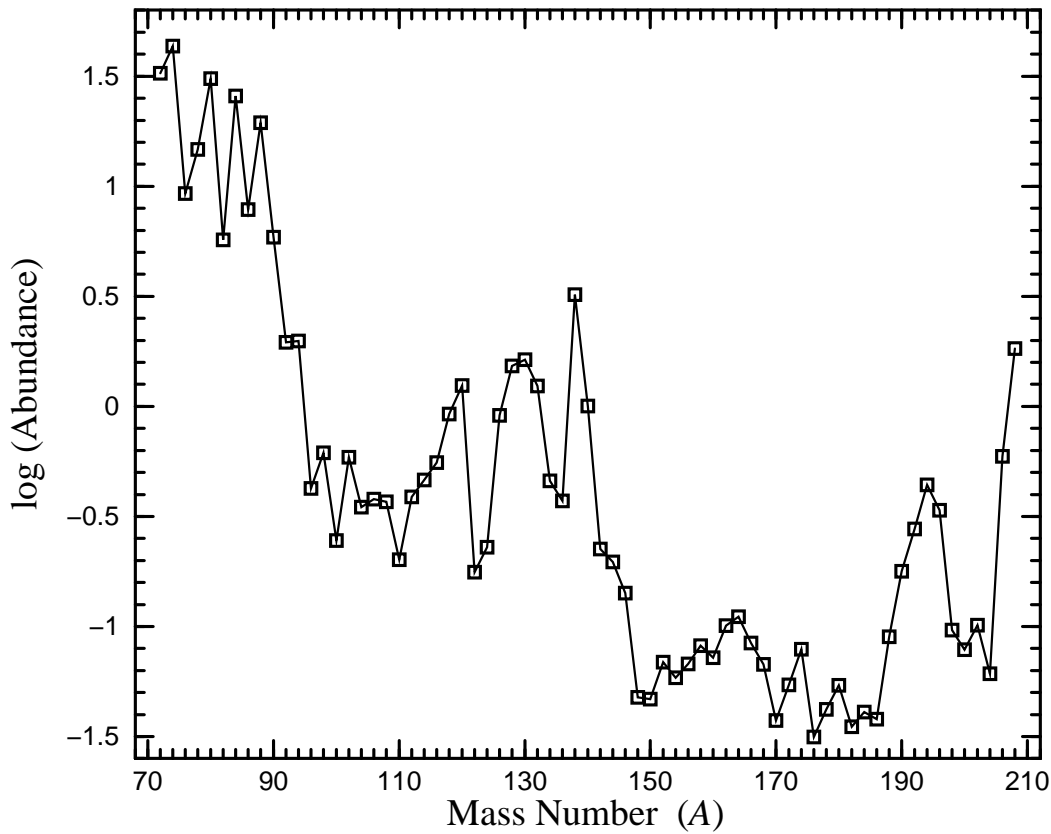,angle=270,scale=0.7}
\end{minipage}
\begin{center}
\begin{minipage}{16.5 cm}
\caption{The logarithm of the solar abundances (normalized so that the
elemental solar abundance of Si is $10^6$) as a function
of mass number $A$. For clarity, only the
abundances of the nuclei with even proton and neutron numbers are shown. 
When there are two or more stable nuclei with the same mass number,
the abundance of the dominant isobar is shown. See \cite{an89}
for the full set of abundances. The peaks at $A=80$, 88, 130, 138,
195, and 208 discussed in the text can be clearly seen (the abundance 
of $^{195}$Pt, which defines the peak at $A=195$,
is essentially the same as the abundance shown for $A=194$). It is
interesting to note that the abundances of $^{118}$Sn and $^{120}$Sn
(with the magic proton number $Z=50$) are comparable to 
those in the peak at $A=130$.}
\end{minipage}
\end{center}
\end{figure}

\subsection{The $s$-process\label{sintro}}
In the $s$-process, neutron capture is slow compared with 
$\beta$ decay so that an unstable nucleus produced by neutron capture 
$\beta$-decays to its stable daughter before it captures another 
neutron. Consequently, the $s$-process proceeds close to the 
$\beta$-stability line. The number of the stable nuclei with mass number 
$A$ produced in an $s$-process event, i.e., the yield $Y_s(A)$, is 
governed by
\be
\dot Y_s(A)=\phi_n[\sigma_{n,\gamma}(A-1)Y_s(A-1)
-\sigma_{n,\gamma}(A)Y_s(A)],
\label{ns}
\ee
where $\dot Y_s(A)$ is the rate of change in $Y_s(A)$, $\phi_n$ is 
the neutron flux, and $\sigma_{n,\gamma}(A)$ is the
neutron-capture cross section (appropriate for the conditions in the 
$s$-process environment) for the stable nucleus with mass number
$A$. When an equilibrium is reached, 
\be
\sigma_{n,\gamma}(A-1)Y_s(A-1)=\sigma_{n,\gamma}(A)Y_s(A),
\ee
which means that the $s$-process yield of a stable nucleus is inversely 
proportional to its neutron-capture cross section. As the stable 
nuclei $^{88}$Sr, $^{138}$Ba, and $^{208}$Pb with the magic neutron 
numbers $N=50$, 82, and 126, respectively, have extremely small 
neutron-capture cross sections, the $s$-process produces abundance 
peaks at these nuclei, which are observed in the solar abundance 
distribution (see Fig. 1).

The $s$-process encounters the magic neutron numbers $N=50$, 82, and 126
at the stable nuclei with proton numbers $Z=38$, 56, and 82, respectively.
As neutron capture is rapid compared with $\beta$ decay during the
$r$-process, this process proceeds on the neutron-rich side of 
the $\beta$-stability line and encounters the same magic neutron 
numbers $N=50$, 82, and 126 at unstable neutron-rich progenitor nuclei 
with smaller proton numbers $Z\sim 30$, 48, and 69 corresponding to 
$A\sim 80$, 130, and 195, respectively. The production of these 
progenitor nuclei is favored by the $r$-process due to their 
relative stability associated with the magic neutron numbers. The 
successive $\beta$ decay of these nuclei following the cessation of 
rapid neutron capture then gives rise to the peaks at $^{80}$Se,
$^{130}$Te, and $^{195}$Pt observed in the solar abundance distribution 
(see Fig. 1).

It is useful to decompose the solar abundances into the separate 
contributions from the $s$-process and the $r$-process, 
respectively. This decomposition is trivial for the nuclei that can 
only be produced by the $s$-process or the $r$-process (i.e., the 
$s$-only or $r$-only nuclei). For example, the stable nucleus 
$^{136}$Ba has a more neutron-rich stable isobar $^{136}$Xe. The 
successive $\beta$ decay of the $r$-process progenitor nucleus with 
$A=136$ stops at $^{136}$Xe and does not contribute to the abundance 
of $^{136}$Ba. Thus, $^{136}$Ba is an $s$-only nucleus. On the other 
hand, the unstable nucleus $^{127}$Te is sandwiched by its two 
stable isotopes $^{126}$Te and $^{128}$Te. The $s$-process follows 
the $\beta$ decay of $^{127}$Te and bypasses $^{128}$Te. Thus, 
$^{128}$Te is an $r$-only nucleus. In addition, as the region 
immediately beyond the heaviest stable nucleus $^{209}$Bi is 
populated by very short-lived nuclei, the long-lived nuclei beyond 
$^{209}$Bi such as $^{232}$Th, $^{235}$U, and $^{238}$U cannot be
produced by the $s$-process and must be attributed to the 
$r$-process. Except for the $s$-only and $r$-only nuclei, a heavy 
nucleus generally receives contributions from both the $s$-process 
and the $r$-process. The decomposition of the solar abundances into 
the solar $s$-process and $r$-process abundances, respectively, can 
be accomplished by a phenomenological approach that is based on 
mostly nuclear physics and observational data or by a full approach 
that is based on nuclear physics, stellar physics, and Galactic 
chemical evolution. In both approaches, the goal is to calculate the 
solar $s$-process abundances directly from theory and then to obtain 
the solar $r$-process abundances by subtracting the $s$-process 
contributions from the total solar abundances. 

The main nuclear physical input for the $s$-process is the 
neutron-capture cross sections. The astrophysical input is the 
intensity and the duration of the neutron flux in individual $s$-process 
events, which can be determined from stellar physics and Galactic 
chemical evolution in the full approach. In the phenomenological 
approach, the astrophysical input is parameterized as a distribution
of the integrated exposure to the neutron flux. A given 
neutron-exposure distribution specifies a function
\be
f_s(A)=\sigma_{n,\gamma}(A)N_s(A),
\label{fs}
\ee
where $N_s(A)$ is the $s$-process abundance of the stable nucleus 
with mass number $A$ that is produced by this neutron-exposure 
distribution. The neutron-capture cross sections and the solar 
abundances of the $s$-only nuclei in different mass regions can be 
used to select the neutron-exposure distribution that gives the 
function $f_{\odot,s}(A)$ as defined in Eq. (\ref{fs}) but for the 
solar $s$-process abundance $N_{\odot,s}(A)$. The solar $r$-process 
abundance $N_{\odot,r}(A)$ can then be obtained as
\be
N_{\odot,r}(A)=N_\odot(A)-N_{\odot,s}(A)=N_\odot(A)-
[f_{\odot,s}(A)/\sigma_{n,\gamma}(A)],
\ee
where $N_\odot(A)$ is the total solar abundance of the stable 
nucleus with mass number $A$. Figure 2a shows the solar $r$-process 
abundances calculated from this phenomenological approach by 
Arlandini et al. \cite{ar99}.

\begin{figure}
\begin{minipage}{2 cm}
\makebox[1cm]{}
\end{minipage}
\begin{minipage}{8 cm}
\epsfig{file=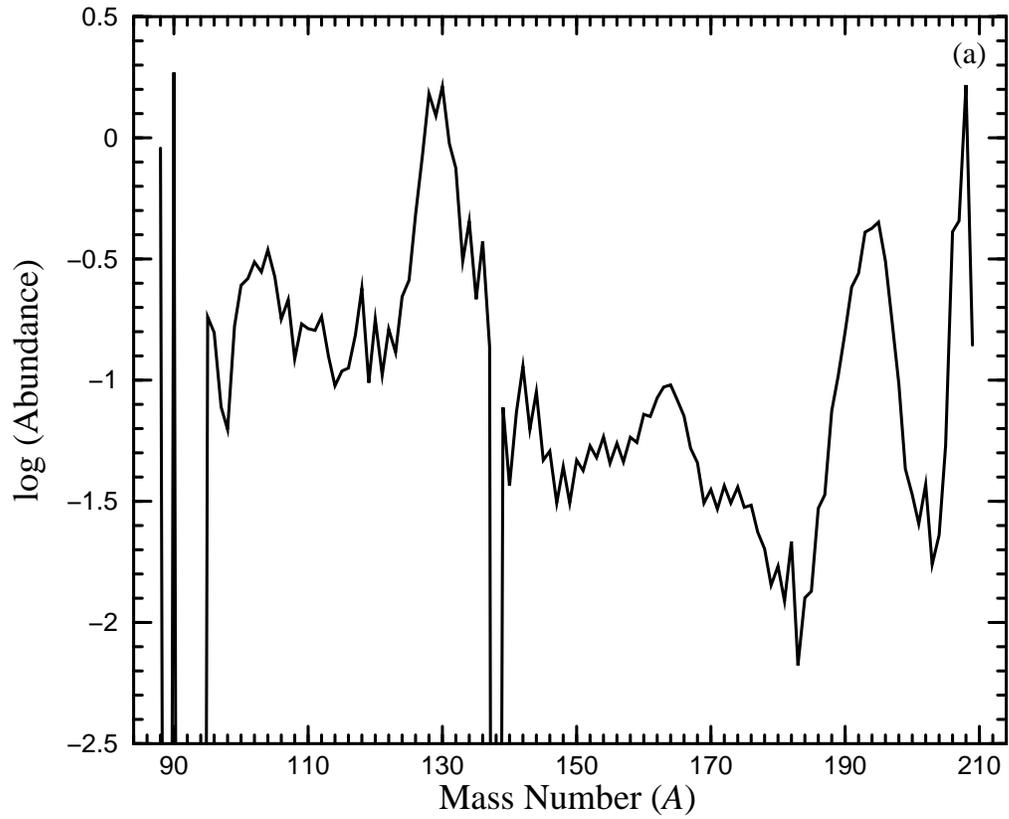,angle=270,scale=0.68}
\epsfig{file=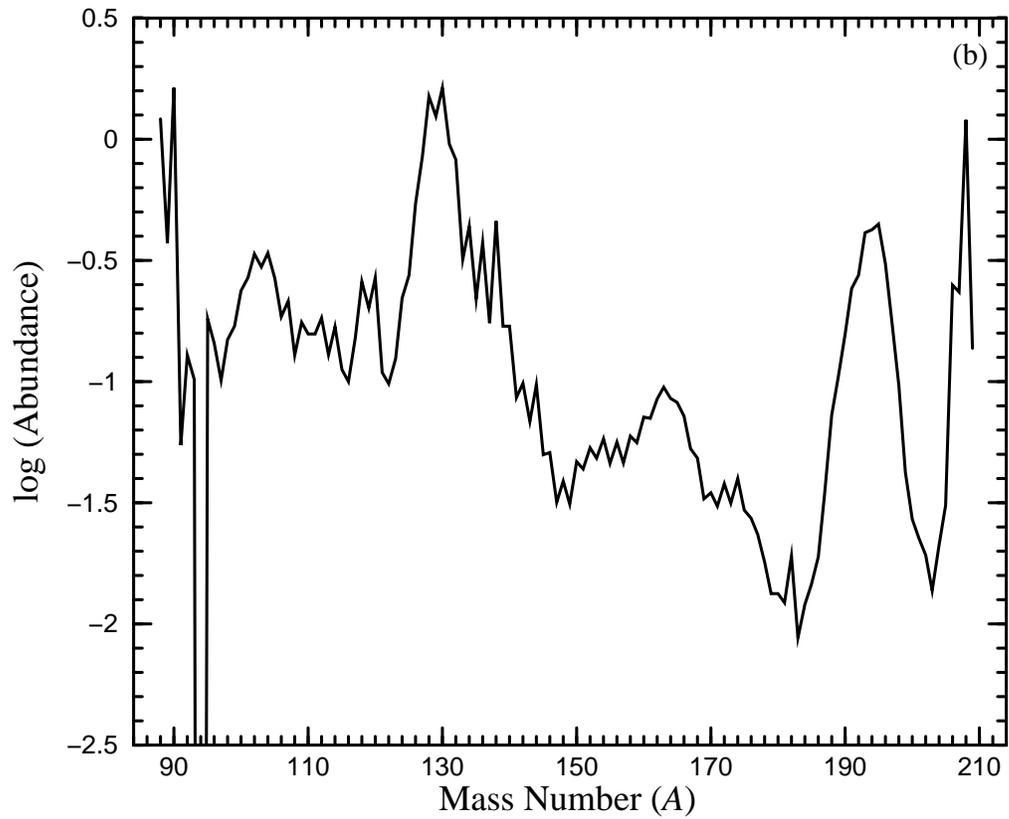,angle=270,scale=0.68}
\end{minipage}
\begin{center}
\begin{minipage}{16.5 cm}
\caption{The logarithm of the solar $r$-process abundances (normalized 
so that the elemental solar abundance of Si is $10^6$) as a 
function of mass number $A$. These abundances are obtained by subtracting
the $s$-process contributions calculated from (a) the phenomenological 
approach and (b) models of two AGB stars. See \cite{ar99} for details.}
\end{minipage}
\end{center}
\end{figure}

Clearly, the solar $r$-process abundances obtained above are subject
to possibly large uncertainties when the solar $s$-process fraction
$\beta_{\odot,s}(A)=N_{\odot,s}(A)/N_\odot(A)$ is close
to unity. For example, the phenomenological approach gives
$\beta_{\odot,s}(^{139}{\rm La})=0.83$ while stellar-model
calculations (see below) give $\beta_{\odot,s}(^{139}{\rm La})=0.62$ 
\cite{ar99}. This difference indicates that the uncertainty in the 
solar $r$-process abundance of $^{139}$La may be as large as 
a factor of 2.2. An important issue is 
then the accuracy of the solar $s$-process abundances as calculated
from the phenomenological approach. The systematic error of this
approach can only be assessed by the full approach that calculates 
and incorporates the $s$-process yields in different 
astrophysical environments over the Galactic history prior to the 
formation of the solar system. The phenomenological approach has
identified a weak $s$-process, which produces the nuclei up to 
$^{88}$Sr, and a main $s$-process, which produces the nuclei 
$^{88}$Sr and above. The weak $s$-process occurs in massive stars
while the main $s$-process occurs in low-mass asymptotic giant 
branch (AGB) stars (e.g., \cite{ka89}). Arlandini et al. 
\cite{ar99} calculated the $s$-process yields of two AGB stars with a 
metallicity of half the solar value but with masses of $1.5\,M_\odot$ 
and $3\,M_\odot$ ($M_\odot$ being the mass of the sun), respectively. 
It is encouraging that the main 
component of the solar $s$-process abundances as calculated from the
phenomenological approach can be essentially reproduced by averaging
the $s$-process yields of these two stars (see Fig. 2b). However, 
this should be viewed only as a significant step towards checking 
and correcting the phenomenological results by the full approach.
In using the solar $r$-process abundances obtained by subtracting the 
$s$-process contributions as calculated from the phenomenological 
approach or a few sample stellar models, it is important to
recognize that these results are generally reliable when 
$\beta_{\odot,s}(A)$ is small, but may have substantial
uncertainties when $\beta_{\odot,s}(A)$ is close to unity. In addition,
the solar $r$-process abundances at $A=206$--209 shown in Fig. 2 may
have to be significantly revised due to important contributions from
low-metallicity AGB stars \cite{ga98}.

Another potential problem with the solar $r$-process abundances obtained 
by subtracting the $s$-process contributions is that some nuclei may be
produced by processes other than the $s$-process and the $r$-process.
The small contributions from the so-called ``$p$-process'' were considered
in \cite{ka89}. However, it is rather difficult to estimate the 
contributions to the solar abundances of the nuclei above the Fe group but
with $A\lesssim 90$ that may come from the $\alpha$-process to be 
discussed in \S\ref{alpha}. In view of this difficulty, the solar 
``$r$-process'' abundances at $A<88$ with a peak at $A=80$ (possibly
corresponding to the progenitor nuclei with $N=50$) will not be discussed 
here.

\subsection{The $r$-process\label{rintro}}
The pattern of the solar $r$-process abundances (the solar 
$r$-pattern) obtained in connection with the $s$-process studies has
played an essential role in the understanding of the $r$-process. As 
mentioned in \S\ref{sintro}, the peaks at $A=130$ and 195 in this pattern
can be accounted for by the nuclear properties associated with the
magic neutron numbers $N=82$ and 126, respectively. It is convenient
to consider the nuclear physical aspects of the $r$-process 
separately from the astrophysical aspects that treat the conditions
in the $r$-process environment. Provided that an $r$-process occurs,
the resulting yield pattern is mostly determined by the nuclear 
systematics of the interplay between neutron capture, 
photo-disintegration, $\beta$ decay, and possibly fission. For the nuclei
with proton numbers $Z<80$, fission may be ignored. Then the number of the
nuclei $(Z,A)$ with proton number $Z$ and mass number $A$ produced in an
$r$-process event, i.e., the yield $Y(Z,A)$, is governed by
\bea
\dot Y(Z,A)&=&n_n\langle v\sigma_{n,\gamma}(Z,A-1)\rangle Y(Z,A-1)
+\lambda_{\gamma,n}(Z,A+1)Y(Z,A+1)+\lambda_{\beta 0}(Z-1,A)Y(Z-1,A)\nonumber\\
&&+\lambda_{\beta 1}(Z-1,A+1)Y(Z-1,A+1)
+\lambda_{\beta 2}(Z-1,A+2)Y(Z-1,A+2)\nonumber\\
&&+\lambda_{\beta 3}(Z-1,A+3)Y(Z-1,A+3)
-n_n\langle v\sigma_{n,\gamma}(Z,A)\rangle Y(Z,A)-
\lambda_{\gamma,n}(Z,A)Y(Z,A)\nonumber\\
&&-[\lambda_{\beta 0}(Z,A)+\lambda_{\beta 1}(Z,A)+\lambda_{\beta 2}(Z,A)
+\lambda_{\beta 3}(Z,A)]Y(Z,A),
\label{rnw}
\eea
where $n_n$ is the neutron number density, 
$n_n\langle v\sigma_{n,\gamma}(Z,A)\rangle$ is the thermally averaged 
neutron-capture rate, $\lambda_{\gamma,n}(Z,A)$ is the photo-disintegration 
rate, and $\lambda_{\beta 0}(Z,A)$, $\lambda_{\beta 1}(Z,A)$,
$\lambda_{\beta 2}(Z,A)$, and $\lambda_{\beta 3}(Z,A)$ are the rates for
$\beta$ decay followed by emission of 0, 1, 2, and 3 neutrons,
respectively. Equation (\ref{rnw}) and the like form an $r$-process 
reaction network. Some approximations that are either helpful in 
understanding the results from full network calculations or used
in place of such calculations are discussed below.

\subsubsection{$(n,\gamma)\protect\rightleftharpoons(\gamma,n)$ 
equilibrium\label{nggn}}
When both neutron capture and 
photo-disintegration occur much faster than $\beta$ decay, a
statistical $(n,\gamma)\rightleftharpoons(\gamma,n)$ equilibrium
is achieved and the yields $Y(Z,A)$ and $Y(Z,A+1)$ of two
neighboring isotopes satisfy
\be
{Y(Z,A+1)\over Y(Z,A)}={n_n\langle v\sigma_{n,\gamma}(Z,A)\rangle
\over\lambda_{\gamma,n}(Z,A+1)}=
n_n\left({2\pi\hbar^2\over m_ukT}\right)^{3/2}
\left({A+1\over A}\right)^{3/2}{G(Z,A+1)\over 2G(Z,A)}
\exp\left[{S_n(Z,A+1)\over kT}\right],
\label{nge}
\ee
where $T$ is the temperature, $\hbar$ is the Planck constant, 
$k$ is the Boltzmann constant, $m_u$ is the atomic mass unit, $G(Z,A)$ is 
the nuclear partition function, and $S_n(Z,A+1)$ is the neutron separation
energy. The most abundant isotope in the isotopic chain 
of $Z$ has a neutron separation energy of
\bea
S_n^0&\approx&kT\ln\left[{2\over n_n}
\left({m_ukT\over 2\pi\hbar^2}\right)^{3/2}\right]\nonumber\\
&=&\left({T\over 10^9\ {\rm K}}\right)\left\{2.79+0.198
\left[\log\left({10^{20}\ {\rm cm}^{-3}\over n_n}\right)
+{3\over 2}\log\left({T\over 10^9\ {\rm K}}\right)\right]\right\}\ {\rm MeV},
\label{sn}
\eea
which can be seen from Eq. (\ref{nge}) by setting 
$Y(Z,A+1)\approx Y(Z,A)$ and neglecting the relatively small 
differences in the nuclear partition function and the mass number.
As $S_n^0$ only depends on $n_n$ and $T$, the most abundant isotopes 
in different isotopic chains have approximately the same neutron 
separation energy, which is $\sim 2$--3 MeV for typical conditions
during the $r$-process. Due to the odd-even effect caused by pairing, the 
most abundant isotope always has an even $N$. For this reason, it 
may be more appropriate to characterize the most abundant isotope
in the isotopic chain of $Z$ by a two-neutron separation energy 
$S_{2n}(Z,A+2)=S_n(Z,A+2)+S_n(Z,A+1)\approx 2S_n^0$ \cite{go92}.

The conditions for $(n,\gamma)\rightleftharpoons(\gamma,n)$ equilibrium 
were first
examined in \cite{ca83b} based on steady-flow calculations 
(see \S\ref{sflo}) and
were found to be $T\gtrsim 2\times 10^9$ K and 
$n_n\gtrsim 10^{20}$ cm$^{-3}$. Later studies \cite{bo96,go96}
emphasized the effects of the nuclear physical input, such as 
the masses of neutron-rich nuclei far from stability, on the
conditions for $(n,\gamma)\rightleftharpoons(\gamma,n)$ equilibrium.
Essentially all of the nuclear physical input for the 
$r$-process is calculated from theory (see \cite{co91} for a review)
and different calculations sometimes give very different results (e.g., 
\cite{go92,bo96,go96}). The uncertainties in the theoretical results can
be reduced by making use of the existing but limited experimental 
knowledge on neutron-rich nuclei far from stability (e.g., \cite{pf01}).
Based on full network calculations with two 
different sets of nuclear physical input, it was found that 
$(n,\gamma)\rightleftharpoons(\gamma,n)$ equilibrium is obtained
for $n_n\gtrsim 10^{20}$ cm$^{-3}$ at
$T=2\times 10^9$ K and for $n_n\gtrsim 10^{28}$ cm$^{-3}$ at 
$T=10^9$ K \cite{go96}. In general, the higher the temperature
and the neutron number density are, the better
$(n,\gamma)\rightleftharpoons(\gamma,n)$ equilibrium holds. 

\subsubsection{steady flow and steady $\beta$-flow\label{sflo}}
In a steady flow, $\dot Y(Z,A)=0$ for all the nuclei in the reaction
network and the yields of these nuclei are determined by a system of
linear algebraic equations. Sample steady-flow $r$-process calculations
can be found in \cite{ca83a}. The steady-flow solution also satisfies
\be
\lambda_\beta(Z-1)Y(Z-1)=\lambda_\beta(Z)Y(Z),
\label{beta}
\ee
where $Y(Z)=\sum_AY(Z,A)$
and $\lambda_\beta(Z)=\sum_A\lambda_\beta(Z,A)Y(Z,A)/Y(Z)$ with 
$\lambda_\beta(Z,A)$ being
the total $\beta$-decay rate of the nucleus $(Z,A)$.
A special case of the steady flow arises when the
$r$-process is also in $(n,\gamma)\rightleftharpoons(\gamma,n)$ 
equilibrium. In this
case, essentially all the abundance in an isotopic chain is carried 
by one or two isotopes with $S_{2n}\sim2S_{n}^0$ (see \S\ref{nggn}). 
Such isotopes are
called the waiting-point nuclei as the $r$-process must wait for 
their $\beta$ decay in order to produce the nuclei in
the next isotopic chain. The above special case of 
the steady flow is called the steady $\beta$-flow. As can be seen from
Eq. (\ref{beta}),
the yield of a waiting-point nucleus in a steady $\beta$-flow
is inversely proportional to its $\beta$-decay rate.

A true steady flow requires that
nuclei be fed into the reaction network from below for a sufficiently 
long time. The results from a steady-flow calculation are valid only 
for certain regions of the network where this condition is met.
It was found that a steady $\beta$-flow may be obtained for the 
waiting-point nuclei between two magic neutron numbers, e.g., those
with $50<N\leq 82$ \cite{kr93}. If steady $\beta$-flows are realized 
in the $r$-process,
then the peaks at $A=130$ and 195 in the solar 
$r$-pattern can be attributed to the extremely small $\beta$-decay 
rates of the waiting-point nuclei with the magic neutron numbers
$N=82$ and 126, respectively.

\subsubsection{fission cycling}
If the $r$-process involves progenitor nuclei with $Z\geq 80$, then
fission should be included in the reaction network. This may lead to
fission cycling, a scenario where the heaviest nucleus produced by
the $r$-process fissions and a cyclic flow occurs between this nucleus 
and its fission fragments in the presence of a large neutron abundance. 
Consider a simple example where $(n,\gamma)\rightleftharpoons(\gamma,n)$ 
equilibrium holds and fission occurs upon the $\beta$ decay
of the heaviest waiting-point nucleus with proton number $Z_f$, which 
produces two fragments with fixed proton numbers $Z_1$ and $Z_2$ 
($Z_2>Z_1$), respectively. A
sufficiently long time after the onset of the $r$-process, the nuclei
with $Z<Z_1$ are depleted to negligible abundances by neutron capture
and $\beta$ decay. Then the yields of the waiting-point nuclei
involved in fission cycling are governed by
\be
\dot Y(Z)=\lambda_\beta(Z-1)Y(Z-1)-\lambda_\beta(Z)Y(Z)
\label{yzfc}
\ee
for $Z_1<Z<Z_2$ and $Z_2<Z\leq Z_f$, and
\bea
\dot Y(Z_1)&=&\lambda_\beta(Z_f)Y(Z_f)-\lambda_\beta(Z_1)Y(Z_1),\\
\dot Y(Z_2)&=&\lambda_\beta(Z_f)Y(Z_f)+\lambda_\beta(Z_2-1)Y(Z_2-1)
-\lambda_\beta(Z_2)Y(Z_2).
\label{yzf}
\eea
With feeding of the nuclear flow at the lower proton numbers $Z_1$ and 
$Z_2$ by fission, a steady state can be obtained after a few 
fission cycles. This results in a rather robust yield pattern
with peaks at $A\sim 130$ and 195 \cite{se65}.

\subsubsection{freeze-out}
In an $r$-process event, the neutron number density and the temperature
decrease with time. Neutron capture ceases to be efficient and the 
$r$-process freezes out when
\be
n_n\langle v\sigma_{n,\gamma}\rangle\tau\sim 1,
\label{fo}
\ee
where $\tau$ is the 
timescale over which $n_n$ decreases significantly. The
freeze-out may involve several stages during which
the approximations discussed above are most likely to break down.
Here full network calculations are
especially important. After the freeze-out, the progenitor nuclei 
successively $\beta$-decay towards stability. In particular,
the progenitor nuclei with magic neutron numbers, which are
in the peaks of the $r$-process yield pattern at the
freeze-out, $\beta$-decay to the stable nuclei with
approximately the same mass numbers but with smaller nonmagic 
neutron numbers. The freeze-out pattern may
differ significantly from the final yield pattern due
to modifications by processes such as $\beta$-delayed neutron 
emission after the freeze-out.

\section{Astrophysical Conditions for the $r$-Process\label{cond}}
The photo-disintegration rate $\lambda_{\gamma,n}(Z,A+1)$ is related
to the neutron-capture rate $n_n\langle v\sigma_{n,\gamma}(Z,A)\rangle$
through detailed balance, which leads to Eq. (\ref{nge}) for
$(n,\gamma)\rightleftharpoons(\gamma,n)$ equilibrium. 
The specification of the rates
for neutron capture and photo-disintegration in the $r$-process network
requires the neutron number density $n_n$ and the temperature $T$.
In traditional calculations (e.g., \cite{go92,bo96,go96,kr93}),
a seed nucleus, usually $^{56}$Fe, is irradiated with neutrons 
at constant values of $n_n$ and $T$ for a time $t_{\rm irr}$. The final
yield pattern is then obtained by assuming an instantaneous freeze-out
and taking into account the modifications of the freeze-out pattern 
due to $\beta$-delayed neutron
emission. Such calculations showed that the solar $r$-pattern must be
accounted for by a superposition of the yield patterns resulting from
distinct sets of $n_n$, $T$, and $t_{\rm irr}$. These sets of parameters 
with typical values of $n_n>10^{20}$ cm$^{-3}$, 
$T\sim 10^9$ K, and $t_{\rm irr}\gtrsim 1$ s
were considered to represent the astrophysical conditions in the 
$r$-process events that occurred over the Galactic history prior to
the formation of the solar system.

In a realistic $r$-process event, $n_n$ and $T$ decrease with time.
The new feature of an $r$-process calculation taking this into 
account can be illustrated by considering the phase during
which $(n,\gamma)\rightleftharpoons(\gamma,n)$ equilibrium holds. 
Due to the evolution
of $n_n$ and $T$, the waiting-point nucleus with mass number 
$A_{\rm WP}$ in the isotopic chain of $Z$ changes with time $t$, which
introduces a time dependence into the effective $\beta$-decay rate
$\lambda_\beta(Z,t)=\sum_A \lambda_\beta(Z,A)Y(Z,A)/Y(Z)\approx
\lambda_\beta[Z,A_{\rm WP}(t)]$. As $\lambda_\beta[Z,A_{\rm WP}(t)]$
tends to be very large at early
times when high values of $n_n$ favor the waiting-point nuclei close to
the neutron-drip line, the duration of an $r$-process can be greatly
reduced. By contrast, the waiting-point
nuclei and the effective $\beta$-decay rates do not change in the 
traditional $r$-process calculations when 
$(n,\gamma)\rightleftharpoons(\gamma,n)$ 
equilibrium is assumed. Thus, the neutron irradiation
time $t_{\rm irr}$ used in
the traditional calculations to obtain the solar $r$-pattern
may be much longer than 
the actual timescale for an $r$-process. It is also clear that the
decrease of $n_n$ and $T$ in an $r$-process event causes
$(n,\gamma)\rightleftharpoons(\gamma,n)$ equilibrium to break down 
eventually and
makes the freeze-out more like a multi-stage process rather 
than a sudden one.

The evolution of $n_n$ in an $r$-process event can be described in
terms of the neutron-to-seed ratio
\be
R_{n/s}(t)=Y_n(t)/Y_s(t_0),
\ee
where $Y_n(t)$ is the neutron abundance at time $t$ and $Y_s(t_0)$
is the initial abundance of seed nuclei. If fission can be
ignored, conservation of mass and the total number of nuclei gives
\bea
Y_n(t)+\sum_{(Z,A)}AY(Z,A)&=&Y_n(t_0)+A_sY_s(t_0),\label{mcons}\\
\sum_{(Z,A)}Y(Z,A)&=&Y_s(t_0),\label{ycons}
\eea
where $A_s$ is the mass number of the seed nucleus and the sums extend over
all the nuclei with $A\geq A_s$. Equations (\ref{mcons}) and (\ref{ycons})
can be rewritten as
\be
\bar A(t)+R_{n/s}(t)=A_s+R_{n/s}(t_0),
\ee
where $\bar A(t)=\sum_{(Z,A)}AY(Z,A)/\sum_{(Z,A)}Y(Z,A)$ is the 
average mass number of the nuclei with $A\geq A_s$ that are in the
$r$-process network at time $t$. In general, when the $r$-process freezes 
out at $t=t_{\rm FO}$, $R_{n/s}(t_{\rm FO})\lesssim 1$ and
$\bar A(t_{\rm FO})\approx A_s+R_{n/s}(t_0)$. Thus, the outcome of
an $r$-process event can be simply estimated from the initial
neutron-to-seed ratio and the mass number of the seed nucleus.

The initial neutron-to-seed ratio $R_{n/s}(t_0)$
not only provides a convenient
means to characterize an $r$-process event but also highlights two 
important issues: where do the seed
nuclei come from and how is $R_{n/s}(t_0)$ determined? 
Observations of metal-poor stars to be discussed in \S\ref{obs}
showed that the $r$-process already occurred in the early history
of the Galaxy. This suggests that an $r$-process event cannot rely on 
some previous astrophysical events to provide the seed nuclei and
must produce the seed nuclei within the event itself. The
production of the seed nuclei and the determination of $R_{n/s}(t_0)$
in an $r$-process event can be illustrated
by considering a rather generic scenario in which neutron-rich
material adiabatically expands from high temperature and density.
The parameters characterizing the expansion that results in 
a successful $r$-process
can be taken as a new measure of the astrophysical conditions for
the $r$-process.

\subsection{Adiabatic expansion from high temperature and density\label{adexp}}
The astrophysical input for a nucleosynthesis calculation includes
the initial state and the time evolution of temperature and density.
For electrically-neutral matter expanding from $T\gtrsim 10^{10}$~K, 
the initial nuclear composition is
\bea
Y_p(0)&=&Y_e,\\
Y_n(0)&=&1-Y_e,
\eea
where $Y_e$ is the net electron number per baryon, or the
electron fraction. Neutron-rich matter has $Y_e<0.5$.
For adiabatic expansion, the thermodynamic evolution
of the matter is governed by
\be
TdS+\sum_{(Z,A)}\mu(Z,A)dY(Z,A)=0,
\label{mu}
\ee
where $S$ is the total entropy per baryon for all the particles in
the matter (nucleons, nuclei, $e^-$, $e^+$, and photons) and
$\mu(Z,A)$ is the chemical potential (including the rest mass)
of the nucleus $(Z,A)$. At $T\gtrsim 6\times 10^9$~K (usually
also corresponding to high density), the forward strong and 
electromagnetic reactions are balanced by their reverse reactions. 
This results in nuclear statistical equilibrium (NSE), for which the 
second term on the left-hand side of Eq. (\ref{mu}) vanishes and the 
adiabatic condition reduces to $S$ being constant. 
After NSE breaks down, some entropy is generated by out-of-equilibrium 
reactions. However, this has little effect on considerations of the 
$r$-process \cite{me97}. So $S$ may be taken as constant during
adiabatic expansion. 

If photons are much more abundant than nucleons and nuclei (radiation
dominance, $S\gg 10$ in units of $k$ per baryon),
\be
S\approx {11\pi^2\over 45}\left({kT\over\hbar c}\right)^3{m_u\over\rho}
=3.34\left({T\over 10^9\ {\rm K}}\right)^3
\left({10^5\ {\rm g\ cm}^{-3}\over\rho}\right)
\label{srad}
\ee
for $T\gtrsim 5\times 10^9$~K when $e^-$ and $e^+$ are also abundant. In 
Eq. (\ref{srad}), $\rho$ is the matter density. The coefficient 3.34 in 
Eq. (\ref{srad}) should be replaced by smaller values for 
$T<5\times 10^9$~K due to the annihilation of $e^-$ and $e^+$
(e.g., the proper coefficient is 1.21 for $T\lesssim 10^9$~K).
On the other hand, if
nucleons and nuclei are much more abundant than photons (matter
dominance, $S\lesssim 10$), $S$ depends on $T$ and $\rho$ as 
$\ln(T^{3/2}/\rho)$. In general, $S$ provides a relation between $T$ and 
$\rho$ during adiabatic expansion. So only the time evolution of $T$ or 
$\rho$ needs to be specified explicitly for a nucleosynthesis calculation. 
One of the parametrizations used in the literature (\cite{ho97}, but see
\cite{me97,fr99} for different choices) is
\be
T(t)=T(0)\exp\left(-{t/\tau_{\rm dyn}}\right),
\label{tdyn}
\ee
where $\tau_{\rm dyn}$ is the dynamic timescale of the expansion.
Thus, a nucleosynthesis calculation can be done for matter adiabatically
expanding from e.g., $T\sim 10^{10}$~K once the parameters $Y_e$, $S$, 
and $\tau_{\rm dyn}$ are given.

\subsection{Statistical equilibrium and nucleosynthesis\label{qse}}
In NSE, the abundance of the nucleus $(Z,A)$ can be obtained from
$\mu(Z,A)=Z\mu_p+(A-Z)\mu_n$ as
\be
Y(Z,A)={G(Z,A)A^{3/2}\over 2^A}
\left({2\pi\hbar^2\over m_ukT}\right)^{3(A-1)/2}
\left({\rho\over m_u}\right)^{A-1}Y_n^{A-Z}Y_p^Z
\exp\left[{B(Z,A)\over kT}\right],
\label{nse}
\ee
where $B(Z,A)$ is the nuclear binding energy of the nucleus $(Z,A)$.
The nuclear composition in NSE does not depend on the reaction rates
and is completely specified by $Y_e$,
$T$, and $\rho$ together with the constraints of charge and mass
conservation:
\bea
Y_e&=&Y_p+\sum_{(Z,A)}ZY(Z,A),\\
1&=&Y_n+Y_p+\sum_{(Z,A)}AY(Z,A),
\eea
where the sums over nuclei exclude the nucleons.
After NSE breaks down at $T\sim 6\times 10^9$~K, the nuclear composition
is governed by the development of quasi-equilibrium (QSE) clusters.
The total abundance in each QSE cluster is specified by
the rates for the nuclear flows into and out of the cluster, 
but the relative 
abundances of the nuclei within the cluster are the same as given by 
Eq. (\ref{nse}) and
again do not depend on the reaction rates \cite{ht96,me98}.
A familiar example of QSE is $(n,\gamma)\rightleftharpoons(\gamma,n)$ 
equilibrium, for 
which each isotopic chain is a QSE cluster.
The total abundance $Y(Z)$ in the isotopic chain of $Z$ is determined by
the $\beta$ decay of the nuclei in the isotopic chains of $Z-1$ and $Z$,
but the abundance ratio $Y(Z,A+1)/Y(Z,A)$ as given in Eq. (\ref{nge})
can be simply obtained from Eq. (\ref{nse}).

At $6\times 10^9\ {\rm K}\gtrsim T\gtrsim 4\times 10^9\ {\rm K}$,
there are two major QSE clusters: the light one involving neutrons, 
protons, and $\alpha$-particles, and the heavy one involving $^{12}$C 
and heavier nuclei \cite{me98}. 
The abundance of $\alpha$-particles in the light 
cluster is given by Eq. (\ref{nse}) while the relative abundances 
of the nuclei within the heavy cluster can be
estimated by using the same equation. The total abundance in the heavy 
cluster is governed by the net flow out of the light one, which
crucially depends on the reactions $3\alpha\to{^{12}{\rm C}}+\gamma$, 
$\alpha+\alpha+n\to{^9{\rm Be}}+\gamma$ followed by 
$\alpha+{^9{\rm Be}}\to{^{12}{\rm C}}+n$, and all the reverse reactions.
The nuclear composition of the heavy cluster freezes out at
$T\sim 4\times 10^{9}$~K. This composition is slightly modified by capture 
of neutrons and $\alpha$-particles down to $T\sim 2\times 10^9$~K,
at which temperature essentially all charged-particle reactions freeze out
due to the Coulomb barrier. 
If there is a sufficiently high neutron abundance at this point,
$(n,\gamma)\rightleftharpoons(\gamma,n)$ equilibrium takes over and an 
$r$-process starts.

\subsection{Determination of the initial neutron-to-seed ratio\label{alpha}}
As neutron-rich matter with $Y_p(0)=Y_e<0.5$ and $Y_n(0)=1-Y_e$
adiabatically expands from $T\sim 10^{10}$~K, its nuclear composition
evolves as follows based on the discussion in \S\ref{qse}. The first
phase of evolution is in NSE, during which essentially all the protons 
are assembled into $\alpha$-particles. When NSE
breaks down at $T\sim 6\times 10^9$~K, the nuclear composition is 
dominated by neutrons and $\alpha$-particles. During the subsequent 
phase of evolution in QSE, heavy nuclei are produced subject to the
influence of the bottle-neck imposed by the three-body reactions
$3\alpha\to{^{12}{\rm C}}+\gamma$ and
$\alpha+\alpha+n\to{^9{\rm Be}}+\gamma$ that
connect the light QSE cluster and the heavy one. The overall neutron 
excess $\eta=1-2Y_e$ is crucial to the composition of the heavy QSE 
cluster. For $0<\eta<0.05$ that is similar to the 
neutron excess $\eta(Z,A)=1-(2Z/A)$ of the individual nuclei in the Fe 
group, these nuclei are the heavy nuclei favored by QSE as they 
have the largest nuclear binding energy per nucleon $B(Z,A)/A$. 
By contrast, QSE favors the heavy nuclei with $Z\sim 35$--40 and 
$A\sim 90$ (close to the magic neutron number $N=50$)
for $\eta>0.05$. These nuclei have much larger values of $\eta(Z,A)$ 
but similar values of $B(Z,A)/A$ compared with the Fe group nuclei. 
The overall $\eta$ required for producing the heavy nuclei with 
$Z\sim 35$--40 and $A\sim 90$ can be significantly smaller than the values 
of $\eta(Z,A)\sim 0.1$--0.2 for these nuclei if the abundance $Y_\alpha$ 
of $\alpha$-particles is large. This is because 
$\eta\approx X_n+\sum_{(Z,A)}\eta(Z,A)X(Z,A)$ with $X_n$ and $X(Z,A)$ being
the mass fractions of neutrons and nuclei, respectively.
The $\alpha$-particles have no neutron excess and their 
presence allows $\eta$ to be small by reducing the mass fraction of 
the heavy nuclei that have significant neutron excess. The composition
of the heavy nuclei freezes out of QSE at $T\sim 4\times 10^9$~K and is 
modified somewhat by capture of neutrons and $\alpha$-particles down to
$T\sim 2\times 10^9$~K. Thereafter, an $r$-process occurs as the heavy 
nuclei with $Z\sim 35$--40 and $A\sim 90$
become the seed nuclei to capture the remaining neutrons. 
The production of heavy nuclei in the presence of significant neutron and 
$\alpha$-particle abundances was first proposed as the precursor to the 
$r$-process by Woosley and Hoffman \cite{wo92}, who referred to this
as the ``$\alpha$-process.'' A thermodynamic approach to understand this
process was given in \cite{me98}.

For clarity in discussing adiabatic expansion of neutron-rich matter,
the beginning of the expansion is denoted as $t=0$, that of the 
$\alpha$-process is denoted as $t_\alpha$, and that of the $r$-process
is denoted as $t_0$. 
The initial neutron-to-seed ratio $R_{n/s}(t_0)$ for the $r$-process
is determined by the abundances of neutrons and heavy nuclei at the
end of the $\alpha$-process, which in turn depend on the net flow into
the heavy QSE cluster from the light one during the $\alpha$-process.
As discussed in \S\ref{qse}, this flow starts with the reactions
building $^{12}$C. 
The physics involved in the determination of $R_{n/s}(t_0)$ is reflected
by the dependence of $R_{n/s}(t_0)$ on $Y_e$, $S$, and 
$\tau_{\rm dyn}$, and can be illustrated by focusing on the reactions
$\alpha+\alpha+n\to{^9{\rm Be}}+\gamma$ and
$\alpha+{^9{\rm Be}}\to{^{12}{\rm C}}+n$. For the most part of the 
$\alpha$-process, the abundance $Y_9$
of $^9$Be is approximately given by Eq. (\ref{nse}). As $Y_p$ is small 
during the $\alpha$-process, it is more convenient to use $Y_\alpha$
[also given by Eq. (\ref{nse})] to estimate $Y_9$ as:
\be
Y_9\approx 8.66\times 10^{-11}Y_\alpha^2Y_n
\left({\rho\over 10^5\ {\rm g\ cm}^{-3}}\right)^2
\left({10^9\ {\rm K}\over T}\right)^3
\exp\left({1.826\times 10^{10}\ {\rm K}\over T}\right).
\label{y9}
\ee
The validity of Eq. (\ref{y9}) can be understood from the fragility
of $^9$Be. It only requires a photon of energy 1.573 MeV for
$\gamma+{^9{\rm Be}}\to\alpha+\alpha+n$ to occur. Over the temperature
range of the $\alpha$-process, there are always some photons
on the high-energy tail of the Bose-Einstein distribution that
are able to maintain the equilibrium between disintegration and formation
of $^9$Be. Similarly, the net flow into the heavy QSE cluster is quite 
small early in the $\alpha$-process when the temperature is sufficiently 
high for many endothermic reverse reactions such as 
$n+{^{12}{\rm C}}\to{^9{\rm Be}}+\alpha$ to occur at significant 
rates. The impediment of the flow to the heavy nuclei
is more severe for higher 
values of $S$ corresponding to more photons per baryon.

The effect of $S$ on the $\alpha$-process
becomes even more obvious after the assemblage of
neutrons and $\alpha$-particles into heavy nuclei begins in earnest.
The rate of assemblage is controlled by the rate 
for the reaction $\alpha+{^9{\rm Be}}\to{^{12}{\rm C}}+n$, which is
$Y_9Y_\alpha(\rho/m_u)\langle v\sigma_{\alpha,n}({^9{\rm Be}})\rangle
\propto\rho^3$ [see Eq. (\ref{y9})]. For radiation-dominated expansion
over a given temperature range, this rate is proportional to $S^{-3}$
[see Eq. (\ref{srad})]. The integrated assemblage 
over the duration of the $\alpha$-process is then proportional to
$\tau_{\rm dyn}/S^3$. Thus, smaller values of $\tau_{\rm dyn}$
or higher values of $S$ lead to less consumption of neutrons and less 
production of seed nuclei, both of which tend to increase $R_{n/s}(t_0)$.
The main dependence of $R_{n/s}(t_0)$ 
on $Y_e$ is through the specification of $Y_n(t_\alpha)$ at the beginning 
of the $\alpha$-process. The nuclear composition at $t_\alpha$ 
is dominated by neutrons and $\alpha$-particles with 
$Y_\alpha(t_\alpha)\sim Y_e/2$ and
$Y_n(t_\alpha) \sim 1-2Y_e$. Clearly, lower values of $Y_e$ give higher
values of $Y_n(t_\alpha)$, which means more neutrons at the end of 
the $\alpha$-process, and hence, higher values of $R_{n/s}(t_0)$.

The combinations of $Y_e$, $S$, and $\tau_{\rm dyn}$ that give rise to
$R_{n/s}(t_0)\approx 100$ in matter adiabatically expanding from 
$T\sim 10^{10}$~K were calculated in \cite{ho97} and are shown in Fig. 3.
For some sets of $S$ and $\tau_{\rm dyn}$, there are two possible values 
of $Y_e$ with one being close to 0.5. This can be explained by the effects
of $Y_e$ on the neutron abundance and on the rate for building heavy nuclei 
during the $\alpha$-process. A low neutron abundance for $Y_e\sim 0.5$ 
gives a low rate for building heavy nuclei due to the low
abundance of $^9$Be [see Eq. (\ref{y9})]. This results in a low abundance 
of the seed nuclei that can compensate for the low neutron abundance
at the end of the $\alpha$-process to give $R_{n/s}(t_0)\approx 100$.
For a fixed $Y_e$, the required values of $S$ and $\tau_{\rm dyn}$ to give
$R_{n/s}(t_0)\approx 100$ approximately follow the scaling 
$S\propto\tau_{\rm dyn}^{1/3}$. This is because the integrated assemblage 
of seed nuclei during the $\alpha$-process depends on $\tau_{\rm dyn}/S^3$.
Consequently, for the smallest value of $\tau_{\rm dyn}=0.0039$~s shown in
Fig. 3, $R_{n/s}(t_0)\approx 100$ can be obtained over a relatively broad
range of $Y_e$ for $S\lesssim 100$. The curves for the three very different
values of $\tau_{\rm dyn}$ shown in Fig. 3 appear to converge in the
region of $S\lesssim 10$ and $Y_e<0.2$. This is because for such small
values of $S$ corresponding to high $\rho$, the bottle-neck imposed by the
three-body reactions in the production of the seed nuclei is no longer
important [see Eq. (\ref{y9})]. The two major candidate astrophysical
environments for the $r$-process to be discussed in \S\ref{mod} are
characterized by $S\sim 100$ (neutrino-driven wind from core-collapse 
supernovae) and $S\lesssim 10$ (ejecta from neutron star mergers), 
respectively.

The combinations of $Y_e$, $S$, and $\tau_{\rm dyn}$ shown in Fig. 3
are considered to represent the conditions for an $r$-process that can 
produce the nuclei with $A\sim 195$ from the seed nuclei with $A_s\sim 90$. 
Similar results were also obtained in \cite{me97,fr99,wi94}. The 
combinations of 
$Y_e$ and $S$ required to produce the nuclei with $A\sim 130$ would lie
to the left of the curves shown in Fig. 3 for fixed values of 
$\tau_{\rm dyn}$. Illustrative $r$-process calculations with different
combinations of $Y_e$, $S$, and $\tau_{\rm dyn}$ were carried out in 
\cite{me97,fr99} and \cite{me92}--\cite{ta94}.

\begin{figure}[tb]
\begin{minipage}{2 cm}
\makebox[1cm]{}
\end{minipage}
\begin{minipage}{8 cm}
\epsfig{file=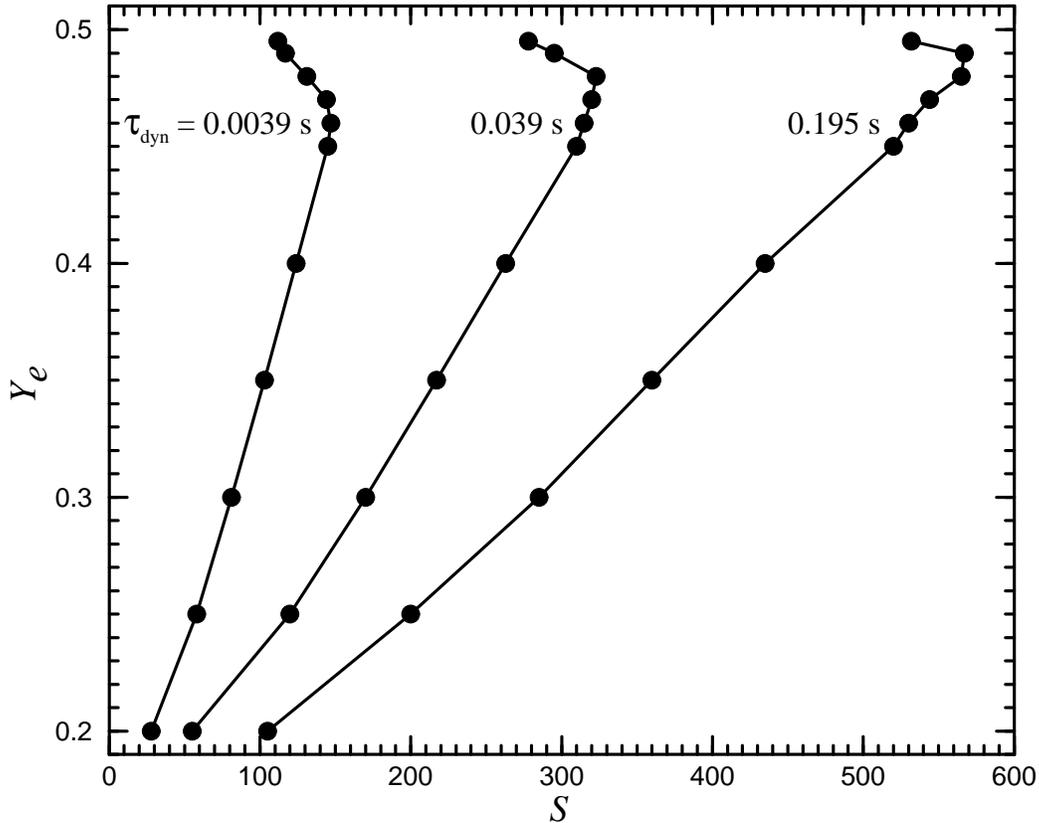,angle=270,scale=0.7}
\end{minipage}
\begin{center}
\begin{minipage}{16.5 cm}
\caption{Combinations of $Y_e$, $S$, and $\tau_{\rm dyn}$
giving rise to an initial neutron-to-seed ratio of 
$R_{n/s}(t_0)\approx 100$ for the $r$-process in
adiabatically expanding matter.
Production of nuclei with $A\sim 195$ is expected. See
\cite{ho97} for details.}
\end{minipage}
\end{center}
\end{figure}

\subsection{Comments on parametric studies}
Both the traditional calculations with constant $n_n$ and $T$ and the more
recent calculations based on adiabatic expansion of neutron-rich matter
from high temperature and density are parametric studies of the
astrophysical conditions for the $r$-process. These studies are largely
independent of specific models and are useful in assessing the potential
of any astrophysical environment to be an $r$-process site. An important
difference between the traditional calculations and the calculations based
on adiabatic expansion lies in the production of the nuclei with
$A\lesssim 90$--110. These nuclei are produced from the seed nucleus
$^{56}$Fe by the $r$-process in the traditional calculations. However,
the seed nuclei for the calculations based on adiabatic expansion are
produced by the $\alpha$-process prior to the $r$-process and usually
already have
$A\sim 90$ (see \S\ref{alpha}). In fact, significant production by the
$\alpha$-process extends to the nuclei with $A\sim 110$ \cite{wo92}.
Thus, $r$-process production truly occurs for $A>110$ in the calculations 
based on adiabatic expansion. These
calculations also shed some new light on the
results from the traditional calculations. As discussed earlier,
the waiting-point nucleus in a given isotopic chain changes due to
the decrease of $n_n$ and $T$ during the $r$-process, which means that
the parameter $t_{\rm irr}$ used in the traditional calculations 
does not represent the actual timescale for an $r$-process. On the
other hand, adiabatic expansion that can produce the peaks at $A=130$ 
and 195 in the solar $r$-pattern was shown to result in
steady $\beta$-flow at the freeze-out of the 
$r$-process with values of $n_n$ and $T$ similar to those obtained from 
the traditional calculations \cite{fr99}. Sample time evolution of $n_n$
for adiabatic expansion with fixed $Y_e$ and $\tau_{\rm dyn}$ but 
different values of $S$ \cite{fr99} is shown in Fig. 4.

\begin{figure}[tb]
\begin{minipage}{2 cm}
\makebox[1cm]{}
\end{minipage}
\begin{minipage}{8 cm}
\epsfig{file=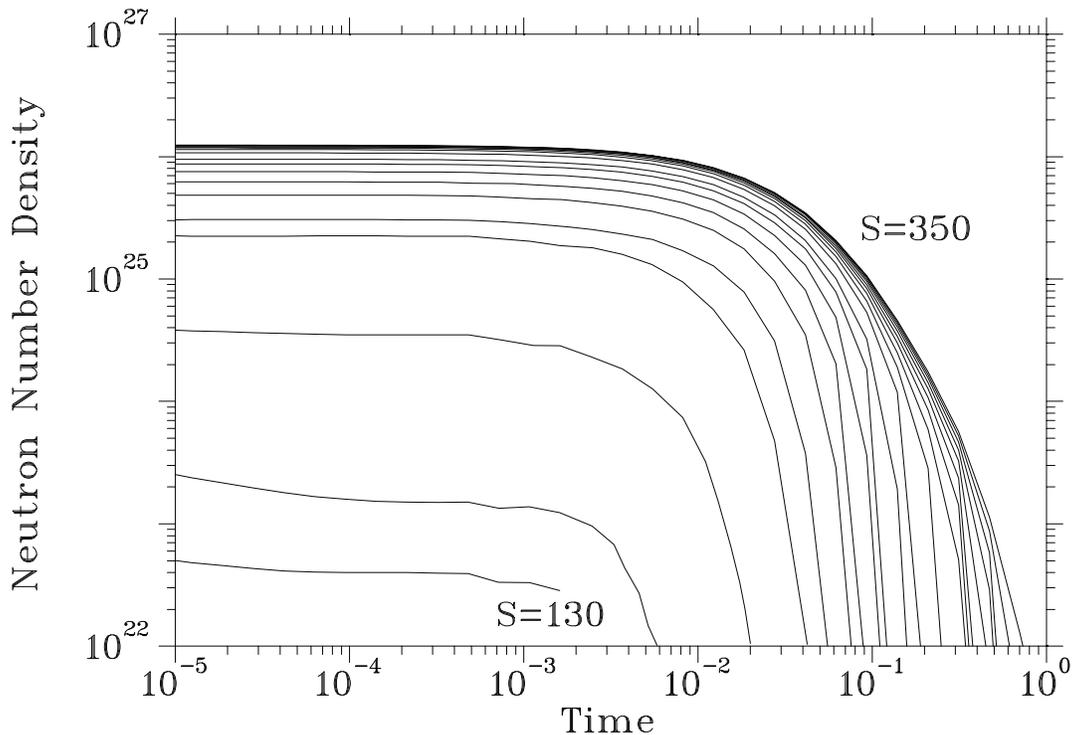,scale=0.57}
\end{minipage}
\begin{center}
\begin{minipage}{16.5 cm}
\caption{Neutron number density $n_n$ as a function of time $t$
for adiabatic expansion with $Y_e=0.45$, $\tau_{\rm dyn}\sim 0.05$ s,
and $S=130$--350. A superposition of these conditions with approximately
equal weights can account for the solar $r$-pattern from the peak at
$A=130$ up to the peak at $A=195$.
Note that the dynamic timescale $\tau_{\rm dyn}$ 
used here corresponds to a parametrization of $T(t)$ 
that is different from Eq. (\ref{tdyn}). See \cite{fr99} for details.}
\end{minipage}
\end{center}
\end{figure}

Some comments on the use of the solar $r$-pattern in parametric studies 
are in order. The traditional calculations require that a
suitable superposition of the yield patterns resulting from different 
combinations of $n_n$, $T$, and $t_{\rm irr}$ match the 
solar $r$-pattern shown in Fig. 2 plus a peak at $A=80$. There are
several potential problems with this requirement. First, the solar 
``$r$-process'' abundances of the nuclei above the Fe group but with 
$A\lesssim 90$--110 were derived by subtracting 
the calculated $s$-process contributions from the total solar abundances
\cite{ka89} and may contain substantial contributions from the
$\alpha$-process, which can produce many of these nuclei \cite{wo92}.
Therefore, the results based on fitting the
solar ``$r$-pattern'' associated with the peak at $A=80$ and possibly
up to the nuclei with $A\sim 110$ must be viewed
with caution. Furthermore, as will be discussed in \S\ref{obs}, the basic 
templates underlying the overall solar $r$-pattern may be more complicated 
than those employed in the traditional calculations. 
The $r$-process yield pattern also depends on the nuclear physical input,
almost all of which are calculated from theory with little experimental
guidance. In view of the above 
issues, parametric studies may be of more general use if they are based on
the simple criterion that
an $r$-process must have an initial neutron-to-seed ratio
$R_{n/s}(t_0)=\langle A\rangle - A_s$ in order to produce nuclei 
with an average mass number $\langle A\rangle$ from the seed nuclei
with mass number $A_s$. In contrast to the case of the $r$-process,
the essential nuclear physical input is known for the
$\alpha$-process that determines $R_{n/s}(t_0)$ and $A_s$.

Parametric studies of the $r$-process have the advantage that no specific
astrophysical models are required. This also leads to some serious flaws.
For example, the parametric studies based on adiabatic expansion cannot 
give the absolute amount of material that has a specific combination of
$Y_e$, $S$, and $\tau_{\rm dyn}$. Furthermore, a real $r$-process
environment may be improperly characterized by some parametric studies.
This flaw has already been demonstrated by the traditional calculations
with constant $n_n$ and $T$. In addition to the problem with the parameter
$t_{\rm irr}$, the freeze-out of the $r$-process is treated as instantaneous
in such calculations. However, the actual freeze-out is a multi-stage
process during which the $r$-process yield pattern may be smoothed
independent of $\beta$-delayed neutron emission 
(e.g., \cite{fr99,ho93}).
It was also proposed \cite{su97} that effects during the freeze-out could
be responsible for the small peak in the rare earth region 
($A\sim 160$) of the solar $r$-pattern (see Fig. 2). 
Such details of the freeze-out are revealed only when the time evolution
of $n_n$ and $T$ in the $r$-process event is taken into account.
In many respects,
the calculations based on adiabatic expansion are more realistic.
However, it is
quite possible that they still miss some important ingredient. For example,
additional parameters characterizing the neutrino fluxes may be required
if the $r$-process occurs in core-collapse supernovae 
(see \S\ref{nueff}).

\section{Astrophysical Models of the $r$-Process\label{mod}}
A neutron-rich environment is required for an $r$-process. It is interesting
to survey the possibilities to obtain such environments during the evolution
of the universe. After the big bang, the universe goes through some
phase transitions, which might produce density inhomogeneities of such
large sizes that only neutrons but not protons can diffuse to uniformity
prior to the onset of big bang nucleosynthesis. This segregation of
neutrons and protons could result in neutron-rich regions where a primordial
$r$-process might occur (e.g., \cite{ap88,ra94}). 
The details of such a scenario will not
be discussed here as its prediction of a minimum $r$-process enrichment
lacks observational support. 
The major outcome of big bang nucleosynthesis is an overall proton-rich
nuclear composition with
$\approx 76\%$ of H and $\approx 24\%$ of $^4$He by mass. This composition
has hardly changed over the history of the universe. Subsequent to big bang 
nucleosynthesis, neutron excess can be produced locally by processes
of weak interaction inside stars. Indeed,
the small neutron excess generated by $\beta^+$ decay
during evolution of stars with masses of $M>10\,M_\odot$
is crucial to the production of the solar abundance pattern for the
nuclei from $^{23}$Na to the Fe group (e.g., \cite{ar73}).
Weak interaction also participates in the reaction
sequence of $p+{^{12}{\rm C}}\to{^{13}{\rm N}}+\gamma$,
${^{13}{\rm N}}\to{^{13}{\rm C}}+e^++\nu_e$, and 
$\alpha+{^{13}{\rm C}}\to{^{16}{\rm O}}+n$, which is
the main neutron source for the $s$-process in AGB stars
(e.g., \cite{ga98}).
The most dramatic production of neutron excess occurs during the 
deleptonization of a protoneutron star that is
produced by the gravitational collapse of a stellar core. Over a period of
$\sim 10$ s, the protoneutron star loses nearly all of its initial
electrons and protons through the neutronization reaction $e^-+p\to n+\nu_e$.
Concurrently with the deleptonization, the protoneutron star also releases
its enormous gravitational binding energy through emission of $\nu_e$,
$\bar\nu_e$, $\nu_\mu$, $\bar\nu_\mu$, $\nu_\tau$, and $\bar\nu_\tau$.
The interaction between the neutrinos and the material above the
protoneutron star drives a supernova explosion on a timescale of $<1$ s 
(e.g., \cite{be85}) and ejects a small amount of material in a wind that
lasts for a period of $\sim 10$ s (e.g., \cite{du86}). 
The neutron-richness of
the wind material is set by the competition between the reactions 
$\nu_e+n\to p+e^-$ and $\bar\nu_e+p\to n+e^+$ (e.g., \cite{qi93,qi96}). 
The high neutron number density of $>10^{20}$ cm$^{-3}$ required for the 
$r$-process (see \S\ref{cond} and Fig. 4) may be obtained in the 
neutrino-driven 
wind associated with protoneutron star formation in core-collapse 
supernovae. Material with such neutron number density may also be ejected
when an old neutron star is disrupted during its merging with 
another neutron star or a black hole. The core-collapse supernova and
neutron star merger models of the $r$-process are discussed in detail 
below.

\subsection{Core-collapse supernova models of the $r$-process\label{snr}}
The evolution, explosion, and nucleosynthesis of massive stars are a rich
subject (see \cite{wo02} for a recent review). Only a simple sketch of the
evolution and explosion of such stars is given
here to provide some general astrophysical context for the discussion of
the $r$-process. Through various stages of hydrostatic burning,
massive stars with $M>10\,M_\odot$ develop an onion-skin structure
with an Fe core surrounded by shells of successively lighter elements from Si 
to H. As no more nuclear binding
energy can be released by burning Fe, the Fe core undergoes gravitational
collapse. When the inner core reaches nuclear density, it
becomes a protoneutron star and bounces. This launches a
shock, which fails to exit the outer core mainly due to energy loss from
dissociation of Fe. Although a successful mechanism remains to be 
demonstrated (see \cite{be90} for a review), it is considered that the shock
is revived by the neutrinos emitted from the protoneutron star and proceeds
to make a Type II supernova (SN II). Massive stars with 
$8\,M_\odot\lesssim M\lesssim 10\,M_\odot$ develop a degenerate O-Ne-Mg
core with insignificant shells \cite{no84}. The core collapse in this case
is triggered by electron capture \cite{no87} and again leads to protoneutron
star formation and an SN II \cite{hi84,ma88}. A bare white dwarf in a binary
may also collapse into a protoneutron star due to mass accretion from its 
binary companion \cite{no91}. The supernovae resulting from accretion-induced
collapse of white dwarfs are sometimes referred to as the 
``silent supernovae'' 
due to the lack of prominent optical display.

Whether formed by Fe core collapse, 
O-Ne-Mg core collapse, or accretion-induced collapse (AIC) of a white dwarf, 
the protoneutron star releases its gravitational binding energy through
neutrino emission, the characteristics of which have been modeled by
numerical transport calculations (e.g., \cite{ma87}--\cite{li01}).
The gravitational binding energy of a protoneutron star with a mass of 
$M_{\rm NS}\sim 1.4\,M_\odot$ and a radius of $R_{\rm NS}\sim 10$ km
is $\sim (3/5)GM_{\rm NS}^2/R_{\rm NS}\sim 3\times10^{53}$ erg, where $G$
is the gravitational constant. 
This energy is emitted approximately equally in each neutrino species 
over the neutrino diffusion timescale. All neutrino species have
neutral-current scattering on nucleons and diffuse out of the protoneutron 
star on a timescale of $\sim 10$ s.
The neutrino luminosities satisfy
\be
L_{\nu_e}\sim L_{\bar\nu_e}\sim L_{\nu_\mu}\approx L_{\bar\nu_\mu}
\approx L_{\nu_\tau}\approx L_{\bar\nu_\tau}
\ee
and have a typical value of $\sim 10^{51}$ erg s$^{-1}$ for each species.
During diffusion, different neutrino species stop exchanging energy with 
matter in different decoupling regions, the temperatures of which 
characterize the neutrino energy spectra emergent from the protoneutron 
star. Energy exchange with matter occurs mainly through neutral-current 
scattering on electrons for $\nu_\mu$, $\bar\nu_\mu$, $\nu_\tau$, and 
$\bar\nu_\tau$, but mainly through the charged-current reactions 
$\nu_e+n\to p+e^-$ and $\bar\nu_e+p\to n+e^+$ with higher efficiency for 
$\nu_e$ and $\bar\nu_e$. In fact, energy exchange is more efficient through 
$\nu_e+n\to p+e^-$ than through $\bar\nu_e+p\to n+e^+$ as there are more 
neutrons than protons inside the protoneutron star. Consequently,
the decoupling from local thermodynamic equilibrium with matter occurs
first for $\nu_\mu$, $\bar\nu_\mu$, $\nu_\tau$, and $\bar\nu_\tau$,
next for $\bar\nu_e$, and last for $\nu_e$. Thus, the average 
energies corresponding to the emergent neutrino spectra satisfy
\be
\langle E_{\nu_\mu}\rangle\approx \langle E_{\bar\nu_\mu}\rangle
\approx \langle E_{\nu_\tau}\rangle\approx \langle E_{\bar\nu_\tau}\rangle
>\langle E_{\bar\nu_e}\rangle >\langle E_{\nu_e}\rangle
\label{nuen}
\ee
with typical values of $\langle E_{\nu_e}\rangle\approx 11$ MeV,
$\langle E_{\bar\nu_e}\rangle\approx 16$ MeV, and 
$\langle E_{\nu_\mu}\rangle\approx 25$ MeV. 

Close to the protoneutron star, the temperature is $T\gtrsim 10^{10}$ K and 
the material is in NSE with its nuclear composition dominated by nucleons.
This material is heated by the reactions $\nu_e+n\to p+e^-$ and 
$\bar\nu_e+p\to n+e^+$ and expands away from the protoneutron star
as a neutrino-driven wind. Above a radius corresponding to 
$T\sim 6\times 10^9$ K, the neutrino reactions become inefficient due
to the ever-decreasing neutrino flux. The wind material then essentially
undergoes adiabatic expansion with fixed values
of $Y_e$, $S$, and $\tau_{\rm dyn}$
as in the parametric $r$-process study described in \S\ref{cond}.
The parameters $Y_e$, $S$, and $\tau_{\rm dyn}$ in the wind are determined 
by the history of neutrino interaction with the wind material at
$T\gtrsim 10^{10}$ K \cite{qi96}. In particular, $Y_e$ is set at sufficiently
low values of $T$ so that the rates $\lambda_{e^+n}$ and $\lambda_{e^-p}$ for 
the reactions $e^++n\to p+\bar\nu_e$ and $e^-+p\to n+\nu_e$ are unimportant
compared with the rates $\lambda_{\nu_en}$ and $\lambda_{\bar\nu_ep}$ for the 
reactions $\nu_e+n\to p+e^-$ and $\bar\nu_e+p\to n+e^+$. This gives an 
estimate of $Y_e$ as \cite{qi93}
\be
Y_e\approx{\lambda_{\nu_en}\over\lambda_{\nu_en}+\lambda_{\bar\nu_ep}}
={1\over 1+(\lambda_{\bar\nu_ep}/\lambda_{\nu_en})}.
\label{yew}
\ee
With $L_{\nu_e}\sim L_{\bar\nu_e}$ and 
$\langle E_{\bar\nu_e}\rangle >\langle E_{\nu_e}\rangle$, it was shown that
$\lambda_{\bar\nu_ep}>\lambda_{\nu_en}$, and hence, $Y_e<0.5$ can be
obtained at least for some part of the period of $\sim 10$ s over which
the neutrino-driven wind occurs \cite{qi96}. This wind was first proposed 
as a site of the $r$-process in \cite{wb92}.
However, only one SN II model gave the adequate values of $Y_e$, $S$, and 
$\tau_{\rm dyn}$ for producing the peaks at $A=130$ and 195 in the solar
$r$-pattern \cite{wo94}. The typical values of $0.45\lesssim Y_e<0.5$, 
$S\sim 100$, and $0.01\ {\rm s}\lesssim\tau_{\rm dyn}\lesssim 0.1$ s 
obtained in another SN II model \cite{wi94} and in analytic and numerical 
studies of the neutrino-driven wind \cite{qi96,th01} 
were unable to give an $r$-process
(see Fig. 3). Parametric studies of $r$-process nucleosynthesis in the
neutrino-driven wind were carried out in \cite{me92}--\cite{ta94}.

\subsubsection{effects of neutrinos\label{nueff}}
As will be discussed in \S\ref{obsimp}, in order to account for the solar 
$r$-process abundances associated with the peaks at $A=130$ and 195, each 
supernova must eject 
$\sim 10^{-6}$--$10^{-5}\,M_\odot$ of $r$-process material.
Although the current neutrino-driven wind models have difficulty in providing 
the $r$-process conditions, the wind naturally ejects 
$\sim 10^{-6}$--$10^{-5}\,M_\odot$ of material over a period of $\sim 1$ s.
This is because the small heating rate due to the weakness of neutrino
interaction permits material to escape from the deep gravitational potential
of the protoneutron star at a typical rate of 
$\sim 10^{-6}$--$10^{-5}\,M_\odot$ s$^{-1}$ \cite{qi96,th01}. 
Indeed, the ability to eject a tiny but interesting amount
of material was recognized as an attractive feature of the 
neutrino-driven wind model of the $r$-process (e.g., \cite{me92}).
On average, a nucleon obtains $\sim 20$ MeV from each interaction with
$\nu_e$ or $\bar\nu_e$. In order to escape from the protoneutron star 
gravitational potential of $GM_{\rm NS}m_u/R_{\rm NS}\sim 200$ MeV, 
a nucleon in the wind must interact with $\nu_e$ and $\bar\nu_e$ for 
$\sim 10$ times. This suggests that the effects of neutrino interaction
may be important even during the essentially adiabatic expansion of the wind 
material. Some of these effects are discussed below in 
connection with $r$-process nucleosynthesis.

At $T\gtrsim 10^{10}$ K, the evolution of $Y_e$ in the wind material is 
governed by
\be
\dot Y_e=(\lambda_{\nu_en}+\lambda_{e^+n})Y_n-
(\lambda_{\bar\nu_ep}+\lambda_{e^-p})Y_p=
\lambda_{\nu_en}+\lambda_{e^+n}-(\lambda_{\nu_en}+\lambda_{\bar\nu_ep}
+\lambda_{e^+n}+\lambda_{e^-p})Y_e,
\label{yeqw}
\ee
where $Y_n=1-Y_e$ and $Y_p=Y_e$ have been used to obtain the second equality.
For small values of $\tau_{\rm dyn}$, $Y_e$ freezes out with a value
given by Eq. (\ref{yew}) when $T$ is so low that $\lambda_{e^+n}$
and $\lambda_{e^-p}$ can be neglected compared with 
$\lambda_{\nu_en}$ and $\lambda_{\bar\nu_ep}$ but
nucleons still dominate the nuclear composition.
However, if neutrino interaction continues to be 
significant when NSE starts to favor
$\alpha$-particles over nucleons at $T<10^{10}$ K, the equation governing the 
evolution of $Y_e$ becomes
\be
\dot Y_e\approx\lambda_{\nu_en}+(\lambda_{\bar\nu_ep}-\lambda_{\nu_en})
{X_\alpha\over 2}-(\lambda_{\bar\nu_ep}+\lambda_{\nu_en})Y_e,
\ee
which is obtained by neglecting $\lambda_{e^+n}$
and $\lambda_{e^-p}$ and using $Y_n=1-Y_e-(X_\alpha/2)$ and 
$Y_p=Y_e-(X_\alpha/2)$, with $X_\alpha$ being the mass fraction of 
$\alpha$-particles, in the first equality of Eq. (\ref{yeqw}).
In the presence of $\alpha$-particles, the evolution of $Y_e$ tries to reach
\be
Y_e\approx{\lambda_{\nu_en}\over\lambda_{\bar\nu_ep}+\lambda_{\nu_en}}
+{\lambda_{\bar\nu_ep}-\lambda_{\nu_en}\over
\lambda_{\bar\nu_ep}+\lambda_{\nu_en}}\left({X_\alpha\over 2}\right),
\ee
which is increased from the value in Eq. (\ref{yew}) for
$\lambda_{\bar\nu_ep}>\lambda_{\nu_en}$. This so-called 
$\alpha$-effect was first discussed in \cite{fu95}. 
Depending on $\tau_{\rm dyn}$, the $\alpha$-effect may 
significantly decrease the initial neutron-to-seed ratio and hinder 
the $r$-process \cite{mm98}. The effects of neutrino interaction with 
nucleons and nuclei on the evolution of $Y_e$ were studied in detail 
in \cite{mc96}.

For large values of $S$ and $\tau_{\rm dyn}$, the $\alpha$-particles present
an additional problem due to their own interaction with neutrinos. In 
particular, $\nu_\mu$, $\bar\nu_\mu$, $\nu_\tau$, and $\bar\nu_\tau$ with
the highest average energy can induce proton spallation on the 
$\alpha$-particles. The daughter nucleus $^3$H from this process provides
a new path to produce the seed nuclei starting with the reactions 
$\alpha+{^3{\rm H}}\to{^7{\rm Li}}+\gamma$ and
$\alpha+{^7{\rm Li}}\to{^{11}{\rm B}}+\gamma$.
As this path does not involve the inefficient three-body reactions to 
burn the $\alpha$-particles (see \S\ref{qse}), the production of the seed
nuclei can be significantly enhanced, which reduces the initial 
neutron-to-seed ratio and again hinders the $r$-process \cite{me95}.

Provided that the above two problems can be avoided and an $r$-process 
occurs in the neutrino-driven wind, neutrino interaction may have some
direct effects on $r$-process nucleosynthesis. First of all, the progenitor 
nuclei encountered during the $r$-process are extremely neutron-rich and have
large cross sections for $\nu_e$ capture (see \cite{hek00,la01} for recent
calculations). In contrast to $\beta$ decay, 
$\nu_e$ capture on the progenitor nuclei are insensitive to the presence of 
magic neutron numbers. Because the abundance peaks at $A=130$ and 195 in the
solar $r$-pattern are usually attributed to the slow $\beta$-decay rates of 
the progenitor nuclei with the magic neutron numbers $N=82$ and 126,
respectively, no such peaks would be produced if the rates for $\nu_e$ 
capture were to dominate those for $\beta$ decay at the freeze-out of the 
$r$-process \cite{fu95}. However, as the
$r$-process proceeds in the wind material, this material is also expanding 
away from the protoneutron star. Consequently, the neutrino flux, and hence,
the neutrino interaction experienced by the wind material decreases 
during the $r$-process. It is possible that $\nu_e$ capture is significant 
in the early phase of the $r$-process 
but becomes negligible compared with $\beta$ decay at the freeze-out 
\cite{qi97,qi98}. In this case, $\nu_e$ capture can accelerate the progress 
from one isotopic chain to the next and reduce the duration of the 
$r$-process as proposed in \cite{na93}. The effects of $\nu_e$ capture
on the $r$-process flow were also discussed in \cite{mf96,mc97}.

Neutrino interaction may also be important during decay towards stability
after the $r$-process freezes out. Both $\nu_e$ capture and neutral-current
reactions with $\nu_\mu$, $\bar\nu_\mu$, $\nu_\tau$, and $\bar\nu_\tau$ 
can greatly excite the progenitor nuclei, which then deexcite through
neutron emission (see \cite{hek00,la01} for recent
calculations) or possibly fission. It was shown that the solar 
$r$-process abundances of the nuclei with $A=124$--126 and $A=183$--187
may be completely accounted for by neutrino-induced neutron emission from 
the progenitor nuclei in the abundance peaks at $A=130$ and 195, respectively 
\cite{qi97,ha97}. This neutrino postprocessing effect is shown for the mass
region near $A=195$ in Fig. 5. This effect also constrains the total exposure 
to neutrinos after the freeze-out as too much
neutrino postprocessing overproduces the nuclei below the 
abundance peaks at $A=130$ 
and 195 \cite{qi97,ha97}. The effect of neutrino-induced fission
during decay towards stability will be discussed in \S\ref{fis} in connection
with the $r$-patterns observed in metal-poor stars.

\begin{figure}[tb]
\begin{minipage}{2 cm}
\makebox[1cm]{}
\end{minipage}
\begin{minipage}{8 cm}
\epsfig{file=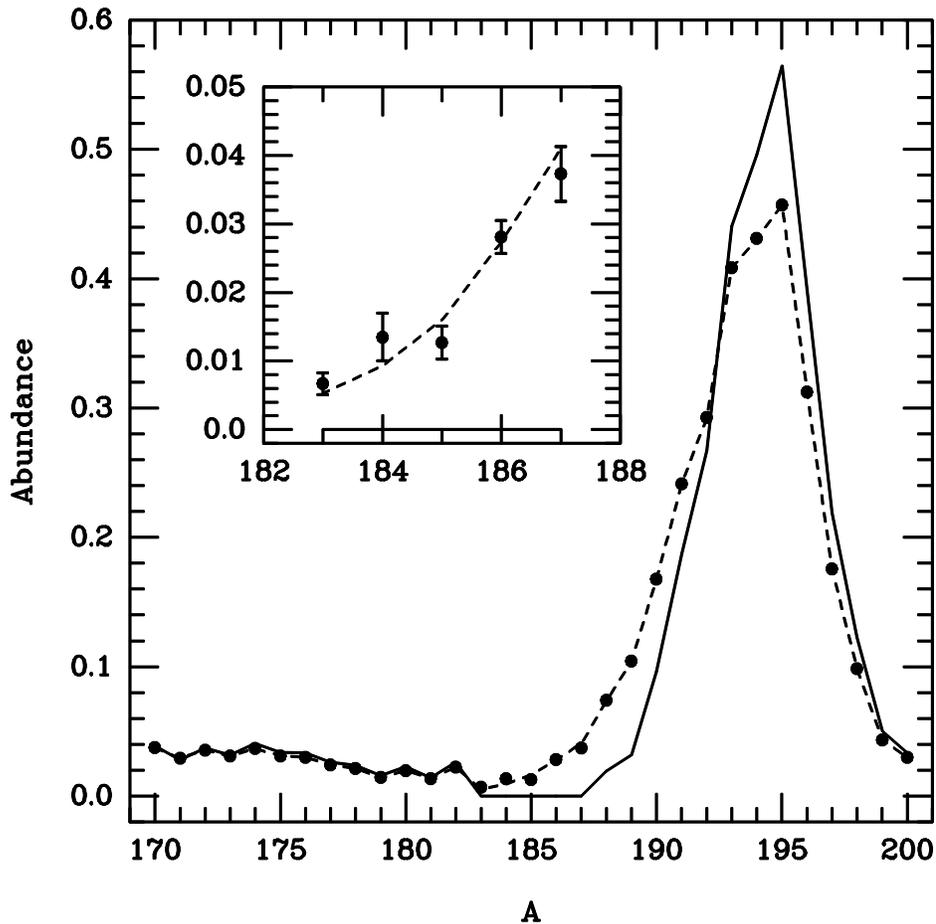,scale=0.7}
\end{minipage}
\begin{center}
\begin{minipage}{16.5 cm}
\caption{Effect of neutrino postprocessing for the mass region
near the abundance peak at $A=195$. 
The abundances before and after the postprocessing are given by
the solid and dashed curves, respectively. The filled circles 
(some with error bars) give the solar
$r$-process abundances. The region with $A=183$--187 is highlighted in the
inset. See \cite{qi97,ha97} for details.}
\end{minipage}
\end{center}
\end{figure}

With a number of possible effects of neutrinos, it is important to do a
self-consistent study that includes all the neutrino interaction processes.
A significant step in this direction was taken in \cite{mm98}, where 
a comparison was made between the results calculated by including neutrino 
interaction to various extent during adiabatic expansion of the wind material.
However, this comparison was based on a fixed set of wind parameters and
the neutrino luminosities used to calculate the neutrino reaction rates 
were not compatible with the parameter $\tau_{\rm dyn}$ adopted for the wind.
In a fully self-consistent study, the wind parameters such as $S$ and 
$\tau_{\rm dyn}$ should be determined by the same neutrino emission 
characteristics that are used to study the neutrino effects on $r$-process
nucleosynthesis. It is essential that such a study 
be carried out in the future. 

\subsubsection{remedies for the neutrino-driven wind model of the 
$r$-process\label{rem}}
As will be discussed in \S\ref{obs}, a number of observations are in support 
of core-collapse supernovae being the major site of the $r$-process. The
amount of $r$-process material required from each supernova to account for
the solar $r$-process abundances associated with the peaks at $A=130$ and 195
is approximately the same as the amount of ejecta in the neutrino-driven
wind. However, current models fail to provide the conditions for an 
$r$-process to occur in the wind. This failure may simply reflect the 
uncertainties in the models and can be remedied when better physical input is 
used. A major uncertainty in the wind models concerns the properties of hot 
and dense matter, which affect the protoneutron star mass $M_{\rm NS}$ and 
radius $R_{\rm NS}$. General relativistic effects are crucial when 
$2GM_{\rm NS}/(R_{\rm NS}c^2)$ ($c$ being the speed of light) approaches
unity. For the nominal values of $M_{\rm NS}=1.4\,M_\odot$ and 
$R_{\rm NS}=10$ km, such effects already lead to significantly larger values 
of $S$ and smaller values of $\tau_{\rm dyn}$ than given by Newtonian 
calculations for the neutrino-driven wind \cite{qi96}. Both increase of 
$S$ and decrease of $\tau_{\rm dyn}$ tend to increase the initial 
neutron-to-seed ratio for the $r$-process (see \S\ref{alpha}). Thus,
massive and/or compact protoneutron stars with large values of
$2GM_{\rm NS}/(R_{\rm NS}c^2)$ may have a wind with adequate conditions for
an $r$-process (see \cite{qi96,th01} and \cite{ca97}--\cite{wa01}). 
The properties of hot and
dense matter also affect neutrino interaction processes in protoneutron 
stars, and hence, the neutrino emission characteristics obtained from
transport calculations (e.g., \cite{bu98}--\cite{po99}). The parameter
$Y_e$ for the neutrino-driven wind is sensitive to the difference 
between the rates $\lambda_{\nu_en}$ and $\lambda_{\bar\nu_ep}$ 
(see Eq. [\ref{yew}]), which are determined by the emission characteristics 
of $\nu_e$ and $\bar\nu_e$ \cite{qi93,qi96,ho99}. It remains to be seen if
low values of $Y_e$ favorable for an $r$-process might result from
neutrino transport calculations based on better understanding of the 
properties of hot and dense matter.

Another major uncertainty in the neutrino-driven wind models
concerns the properties of neutrinos. As mentioned above, $Y_e$ for the wind 
is sensitive to the difference between the rates $\lambda_{\nu_en}$ and 
$\lambda_{\bar\nu_ep}$.
Because the average energy of $\nu_\mu$, $\bar\nu_\mu$, $\nu_\tau$, and 
$\bar\nu_\tau$ is higher than that of either $\nu_e$ or $\bar\nu_e$
[see Eq. (\ref{nuen})], neutrino flavor transformation between the former
group and the latter can greatly affect $\lambda_{\nu_en}$ and 
$\lambda_{\bar\nu_ep}$, and hence, $Y_e$ \cite{qi93}. 
There might also exist a sterile
neutrino that has no standard weak interaction. Flavor transformation
involving the sterile neutrino can dramatically decrease $Y_e$ in the 
neutrino-driven wind by reducing the $\nu_e$ flux \cite{mc99,ca00}. 
With the $\nu_e$ flux greatly reduced in the region where $\alpha$-particles
are being formed, the $\alpha$-effect is also rendered
inoperative \cite{mc99,ca00}. Therefore, sterile neutrinos may play
an essential role in enabling $r$-process nucleosynthesis in 
the neutrino-driven wind. Evidence for the existence of such neutrinos
may come from future experiments on neutrino oscillations.

In typical neutrino-driven wind models, various neutrino interaction
processes provide the energy for driving the wind. The radial
distribution of these energy sources has significant effects on
the parameters $S$ and $\tau_{\rm dyn}$ for the wind \cite{qi96}.
A broad distribution above the radius at which the total energy
deposition rate peaks tends to favor an $r$-process by increasing $S$ 
and decreasing $\tau_{\rm dyn}$ \cite{qi96,th01}.
Such a distribution may be obtained when neutrino
flavor transformation or especially some extra energy source other 
than neutrino interaction (e.g., magnetic field and rotation of the
neutron star \cite{qi96}) is included in the wind models.

In considering the neutrino-driven wind model of the $r$-process,
it is also important to note the difference between the winds associated 
with Fe core collapse, O-Ne-Mg core collapse, and AIC of a white dwarf. 
Although the mass ejection rate and $S$ in the wind are determined near 
the protoneutron star, the structure of the wind further
away may be affected by the outer boundary imposed by the supernova
(e.g., \cite{qi96}). For example, the reverse shock in the 
Fe-core-collapse events may slow down the wind significantly 
\cite{qi96,th01} whereas no such shock is present in the 
O-Ne-Mg-core-collapse or AIC events. This may 
have important consequences for $r$-process nucleosynthesis 
associated with core-collapse supernovae.

\subsubsection{other core-collapse supernova models of the $r$-process}
The neutrino-driven wind model of the $r$-process is closely related
to the neutrino-driven mechanism of core-collapse supernovae.
It is recognized that this mechanism crucially depends on neutrino 
transport and convection (see \cite{be90} for a review), 
a simultaneous and full treatment of which is 
very difficult to implement numerically. Thus, the robustness of this
mechanism remains to be demonstrated. Two alternative explosion
mechanisms were proposed for core-collapse supernovae and both have
associated $r$-process scenarios. In the prompt mechanism, the bounce 
of the inner core upon reaching nuclear density launches
an energetic shock, which then drives a successful explosion
(e.g., \cite{be90}). This mechanism might possibly work for the 
accretion-induced collapse of white dwarfs and the collapse of
O-Ne-Mg cores and maybe some low-mass Fe cores. 
The inner ejecta from a prompt explosion is neutron-rich 
and has been studied as a possible site of the
$r$-process \cite{hi84,wh98,su01}. However, the conditions in this 
ejecta and the amount of ejected $r$-process material are quite 
uncertain as the prompt mechanism underlying this $r$-process model 
is even more problematic than the neutrino-driven mechanism.
Another supernova mechanism relies on the formation of 
magnetohydrodynamic jets following the core collapse (e.g., 
\cite{le70}--\cite{ma01}). While $r$-process nucleosynthesis associated 
with jets was discussed quite some time ago \cite{sy85}, an intriguing 
new scenario was proposed by Cameron \cite{ca01} recently. In this 
scenario, jets are associated with the formation of an accretion disk 
around the protoneutron star. The accretion disk has such 
high densities that electrons are degenerate but neutrinos can
escape. The chemical equilibrium among neutrons, protons, nuclei, and 
electrons results in large abundance ratios of neutrons to heavy nuclei
in the disk. The $r$-process was 
considered to occur as material is transported from the disk
to the base of the jets and ejected. Clearly, more quantitative studies 
of this $r$-process model are worth pursuing.

\subsection{Neutron star merger models of the $r$-process}
Two binary neutron star (NS-NS) systems, PSR 1913+16 \cite{ta89}
and PSR 1534+12 \cite{wo91}, were observed in the Galaxy. The neutron
stars in an NS-NS binary eventually merge due to orbital decay caused by
gravitational radiation. The total time from birth to merger is 
$\approx 4\times 10^8$ yr for 
PSR 1913+16 and $\approx 3\times 10^9$ yr for PSR 1534+12 \cite{ph91}.
Estimates for the rate of NS-NS mergers in the Galaxy range from
$\sim 10^{-6}$ to $\sim 3\times 10^{-4}$ yr$^{-1}$ with the best guess
being $\sim 10^{-5}$ yr$^{-1}$ (e.g., \cite{ph91}--\cite{be02}).
The birth rates of neutron star-black hole (NS-BH) and NS-NS binaries 
are comparable. However, the fraction of NS-BH binaries having the 
appropriate orbital periods for merging within the age of
the universe ($\sim 10^{10}$ yr) is uncertain due to their complicated 
evolution involving mass exchange \cite{ph91}. 
In any case, the total rate
of neutron star (including NS-NS and NS-BH) mergers in the Galaxy is
perhaps $\sim 10^{-5}$ yr$^{-1}$, which is $\sim 10^3$ times smaller 
than the Galactic rate of SNe II \cite{ca99}. This means that each
merger must eject $\gtrsim 10^{-3}\,M_\odot$ of $r$-process
material if neutron star mergers were solely responsible for the solar 
$r$-process abundances
associated with the peaks at $A=130$ and 195 
($\sim 10^{-6}$--$10^{-5}\,M_\odot$ of $r$-process material is 
required from each event in the case of core-collapse supernovae). 

It was estimated that 
$\lesssim 5\%$ of the original neutron star mass may be ejected 
during tidal disruption of the neutron star in an NS-BH merger
\cite{la74,la76}. Estimates for the amount of cold neutron star matter
ejected during an NS-NS merger range from $\sim 10^{-4}\,M_\odot$
to $\sim 4\times10^{-2}\,M_\odot$ \cite{ru97}--\cite{ru01}.
The ejected cold neutron star matter is decompressed and heated to high 
temperature ($\gtrsim 10^9$ K) by $\beta$ decay and possibly 
fission \cite{me89,frt99}. Further evolution of the decompressed
neutron star matter is similar to the adiabatic 
expansion of neutron-rich matter described in \S\ref{cond} and could
lead to an $r$-process.
The $r$-patterns resulting from decompression of cold neutron star matter 
with $Y_e=0.05$, 0.10, 0.12, and 0.15 \cite{frt99} are shown in Fig. 6. 
For these values of $Y_e$, fission cycling occurs and only the heavy 
$r$-process nuclei with $A>130$ are produced. Neutron star mergers 
over the Galactic history would make significant contributions 
or even completely account for the solar $r$-process abundances at $A>130$ 
(e.g., \cite{la74,la76} and \cite{me89}--\cite{sy82}) if
$\sim 10^{-3}\,M_\odot$ of cold neutron star matter 
with $Y_e\approx 0.1$ could be ejected from each event. The implications
of meteoritic data and stellar observations for this $r$-process
model will be discussed in \S\ref{obs}.

\begin{figure}[tb]
\begin{minipage}{2 cm}
\makebox[1cm]{}
\end{minipage}
\begin{minipage}{8 cm}
\epsfig{file=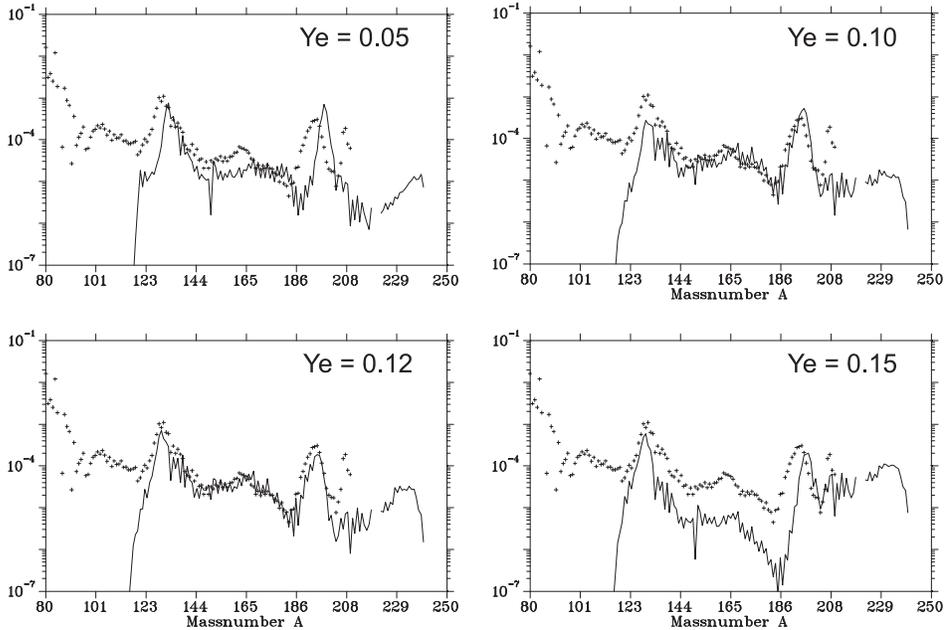,scale=0.7}
\end{minipage}
\begin{center}
\begin{minipage}{16.5 cm}
\caption{The $r$-patterns (solid curves) resulting from decompression of 
cold neutron star matter with $Y_e=0.05$, 0.10, 0.12, and 0.15, which may 
be ejected from neutron star mergers. The solar $r$-pattern (pluses) is
also shown for comparison. See \cite{frt99} for details.}
\end{minipage}
\end{center}
\end{figure}

Several issues regarding the neutron star merger model of the $r$-process
need to be addressed. First of all, numerical calculations 
gave a wide range for the amount of cold neutron star matter ejected 
during an NS-NS merger. It was found that this amount sensitively depends 
on the initial spins of the neutron stars and 
on the equation of state at high 
density \cite{ro99}--\cite{ru01}. General relativity may also be important.
All these issues should be examined by future numerical studies with
higher resolutions (a high-resolution Newtonian calculation was recently
carried out in \cite{ro02}).
Simulations of an NS-BH merger are also needed to determine the actual 
amount of ejecta in such events.
The other major uncertainty in the neutron star merger model of the 
$r$-process concerns $Y_e$ in the ejecta. The values of $Y_e$ not only 
depend on which part of the neutron star is ejected 
but may also be affected
by weak interaction with $e^-$, $e^+$, $\nu_e$, and $\bar\nu_e$.
Although one study indicated that weak interaction has little
effect on $Y_e$ and the ejecta essentially retains the $Y_e$ values of
the cold neutron star matter, such values are still sensitive to the
equation of state at high density \cite{ru01}. In general, as in the
case of core-collapse supernovae, much work with 
improved numerical methods and physical input is required to resolve 
the outstanding issues regarding neutron star mergers as a possible
site of the $r$-process.

\subsection{Similarities between core-collapse supernova and neutron 
star merger models of the $r$-process\label{simi}}
Despite many differences between core-collapse supernovae and neutron 
star mergers, the neutrino-driven wind in the former events and the
cold neutron star matter ejected in the latter events follow similar
thermodynamic evolution, which may lead to the same generic picture for
the $r$-process as described in \S\ref{alpha}. 
Furthermore, a neutrino-driven wind can also occur in
neutron star mergers \cite{ru97}. While $\nu_e$, $\bar\nu_e$, $\nu_\mu$, 
$\bar\nu_\mu$, $\nu_\tau$, and $\bar\nu_\tau$ emitted during such events
have approximately the same energy spectra as in the case of core-collapse
supernovae, the luminosity of $\bar\nu_e$ is $\sim 3$ times larger
than that of $\nu_e$ due to the dominant production of $\bar\nu_e$
via the reaction $e^++n\to p+\bar\nu_e$ \cite{ru97,ru01}. Consequently, 
the neutrino-driven wind from neutron star mergers can have $Y_e$ values
much less than 0.5 [see Eq. (\ref{yew})]. 
It was estimated that $\sim 10^{-4}$--$10^{-2}\,M_\odot$ 
of very neutron-rich material may be ejected in this wind \cite{ru97}.
However, detailed studies of this wind require much better treatment
of neutrino transport than implemented in the existing models \cite{ru01}.
It is important that such studies be carried out in the future.

\section{Observational Studies of the $r$-Process\label{obs}}
Enormous progress has been made in observational studies of the $r$-process
since the last major review by Cowan et al. \cite{co91} in 1991. This 
includes meteoritic evidence for the diversity of $r$-process sources and
demonstration of some regularity in $r$-patterns and of large dispersions
in $r$-process abundances at low metallicities by observations of 
metal-poor stars. These observational advances and their implications for 
the $r$-process are discussed in connection with a three-component model 
for abundances in metal-poor stars. Other models addressing the abundances 
in metal-poor stars and their implications for the $r$-process 
are also described.

\subsection{Meteoritic data and diverse sources for the $r$-process\label{met}}
Meteorites were formed during the birth of the solar system 
$\approx 4.6\times 10^9$ yr ago. They provide important information on
the inventory of radioactive nuclei in the early solar system (ESS).
For example, $^{182}$Hf has a lifetime of $\tau_{182}=1.3\times 10^7$ yr
and $\beta$-decays to $^{182}$Ta, which in turn $\beta$-decays with a
lifetime of 165 days to the stable nucleus $^{182}$W. Any $^{182}$Hf 
that had been incorporated into the meteorites decayed to $^{182}$W long 
ago. Unlike $^{182}$W, the stable nuclei $^{180}$Hf 
and $^{184}$W do not receive any contributions from radioactive decay.
As $^{182}$Hf and $^{180}$Hf are chemically identical, a meteorite with
a high abundance of $^{180}$Hf also has a high initial abundance of
$^{182}$Hf. Therefore, if a significant inventory of $^{182}$Hf existed
in the ESS, there would be a linear correlation between the present
abundance ratios $^{182}$W/$^{184}$W and $^{180}$Hf/$^{184}$W in the
meteorites. Such a correlation was found and gave 
($^{182}$Hf/$^{180}$Hf)$_{\rm ESS}\approx 10^{-4}$ for the 
abundance ratio of $^{182}$Hf to $^{180}$Hf in the ESS \cite{yi02,kl02} 
(see also \cite{le95}--\cite{le00}). 
Similarly, meteoritic measurements gave
($^{129}$I/$^{127}$I)$_{\rm ESS}\approx 10^{-4}$ for the abundance ratio
of $^{129}$I (with a lifetime of $\tau_{129}=2.3\times 10^7$ yr) to
$^{127}$I (stable) in the ESS \cite{re60,br99}.

The nucleus $^{181}$Hf $\beta$-decays with a lifetime of 61 days
in the laboratory and cannot provide a significant branching for neutron 
capture to produce $^{182}$Hf during the
$s$-process. So major production of $^{182}$Hf can only occur in the
$r$-process. Both $^{129}$I and $^{127}$I are essentially pure $r$-process
nuclei. The meteoritic data on $^{182}$Hf and $^{129}$I could not be
explained if there were only a unique $r$-process source with a universal
yield pattern. This can be shown by the method of contradiction.
If $^{182}$Hf and $^{129}$I were produced with
a yield ratio of $Y_{129}/Y_{182}$ by a unique $r$-process source, then
\be
\left({^{129}{\rm I}\over ^{182}{\rm Hf}}\right)_{\rm ESS}>
{Y_{129}\over Y_{182}}.
\label{ihf}
\ee
This is because $^{129}$I survives longer than $^{182}$Hf.
The yield ratio $Y_{129}/Y_{182}$ can be rewritten as
\be
{Y_{129}\over Y_{182}}={Y_{129}\over Y_{127}}
\left({^{127}{\rm I}\over ^{182}{\rm W}_r}\right)_\odot,
\label{iw}
\ee
where the yield ratio $Y_{127}/Y_{182}$ 
has been set to be the same as the solar 
$r$-process abundance ratio of $^{127}$I to $^{182}$W
(with $^{182}$W$_r$ being the part of $^{182}$W 
contributed by the $r$-process) under the assumption of a universal
$r$-process yield pattern.
Equations (\ref{ihf}) and (\ref{iw}) can be rearranged into
\be
{(^{129}{\rm I}/^{127}{\rm I})_{\rm ESS}\over
(^{182}{\rm Hf}/^{180}{\rm Hf})_{\rm ESS}}>{Y_{129}\over Y_{127}}
\left({^{180}{\rm Hf}\over ^{182}{\rm W}_r}\right)_\odot.
\label{ihfw}
\ee
The left-hand side of Eq. (\ref{ihfw}) is $\approx 1$ 
based on the meteoritic data, while the right-hand side is $\approx 4$ 
by taking $Y_{129}/Y_{127}=1.4$ and
$(^{180}{\rm Hf}/^{182}{\rm W}_r)_\odot=0.0541/0.019$ \cite{an89,ar99}.
Thus, Eq. (\ref{ihfw}) is violated by a factor of $\approx 4$ and the
underlying assumption that $^{182}$Hf and $^{129}$I were produced by a
unique $r$-process source must be invalid.

While the diversity of $r$-process sources can be established by using only
the meteoritic data on $^{182}$Hf and $^{129}$I, more may be learned about 
the nature of these sources by making the following assumption. 
Suppose that $^{182}$Hf and $^{129}$I are produced by two distinct kinds
of $r$-process events, which occurred at regular intervals of 
$\Delta_{\rm Hf}$ and $\Delta_{\rm I}$, respectively, over a uniform
production period of $T_{\rm UP}\approx 10^{10}$ yr prior to the formation
of the solar system. If the interval between the last event of each kind
and the formation of the solar system is exactly $\Delta_{\rm Hf}$ or 
$\Delta_{\rm I}$, then
\bea
\left({^{182}{\rm Hf}\over ^{182}{\rm W}_r}\right)_{\rm ESS}&=&
\left({\tau_{182}\over T_{\rm UP}}\right)
{\Delta_{\rm Hf}/\tau_{182}\over\exp(\Delta_{\rm Hf}/\tau_{182})-1},
\label{deltahf}\\
\left({^{129}{\rm I}\over ^{127}{\rm I}}\right)_{\rm ESS}&=&
\left({Y_{129}\over Y_{127}}\right)
\left({\tau_{129}\over T_{\rm UP}}\right)
{\Delta_{\rm I}/\tau_{129}\over\exp(\Delta_{\rm I}/\tau_{129})-1}.
\label{deltai}
\eea
Equations (\ref{deltahf}) and (\ref{deltai}) are valid for
$\Delta_{\rm Hf},\ \Delta_{\rm I}\ll T_{\rm UP}$, in which case the
radioactive nuclei in the ESS were contributed by the last few 
events prior to the formation of the solar system while the stable nuclei
were accumulated over essentially the entire period of $T_{\rm UP}$.
With $(^{182}{\rm Hf}/^{182}{\rm W}_r)_{\rm ESS}=
(^{182}{\rm Hf}/^{180}{\rm Hf})_{\rm ESS}
(^{180}{\rm Hf}/^{182}{\rm W}_r)_\odot=2.8\times 10^{-4}$ and 
$Y_{129}/Y_{127}=1.4$, Eqs. (\ref{deltahf}) and (\ref{deltai}) can be solved 
to give $\Delta_{\rm Hf}=3.3\times 10^7$ yr and 
$\Delta_{\rm I}=1.2\times 10^8$ yr, which should indeed represent two
distinct kinds of $r$-process events. Based on similar arguments, 
Wasserburg et al. \cite{wa96} first pointed out that the meteoritic data on 
$^{182}$Hf and $^{129}$I require diverse sources for the $r$-process
(see also
\cite{qw00}). In general, other nuclei with $A>130$ are produced together 
with $^{182}$Hf while those with $A\lesssim 130$ are produced together with
$^{129}$I. The meteoritic data on $^{182}$Hf and $^{129}$I suggest that
there must be at least two distinct kinds of $r$-process events:
the high-frequency $H$ events producing mainly the heavy $r$-process
nuclei with $A>130$ ($H$ stands for ``high-frequency'' and 
``heavy $r$-process nuclei'') and the low-frequency $L$ events producing 
dominantly the light $r$-process nuclei with $A\lesssim 130$ 
($L$ stands for ``low-frequency'' and ``light $r$-process nuclei'')
\cite{qi98,wa96,qw00}. An average interstellar medium (ISM) is
enriched with the appropriate $r$-process nuclei at a frequency of
$f_H=\Delta_{\rm Hf}^{-1}\sim (3\times 10^7\ {\rm yr})^{-1}$ 
by the $H$ events
and at a frequency of $f_L=\Delta_{\rm I}^{-1}\sim (10^8\ {\rm yr})^{-1}$ 
by the $L$ events.

The frequencies of the $H$ and $L$ events can 
be compared with the frequencies for replenishment of newly-synthesized
material in the ISM by core-collapse supernovae
and neutron star mergers. The total kinetic energy of the ejecta from
a supernova or neutron star merger is typically $\sim 10^{51}$ erg.
The amount of ISM required to dissipate the energy and the
momentum of the ejecta is $M_{\rm mix}\sim 3\times 10^4\,M_\odot$
(e.g., \cite{th98}). For a present event rate of $f_{\rm G}$ in the
Galaxy corresponding to a total gas mass of 
$M_{\rm gas}\sim 10^{10}\,M_\odot$, the frequency for replenishment of 
newly-synthesized material in the ISM by the relevant events is
\be
f_{\rm mix}=(10^7\ {\rm yr})^{-1}
\left[{f_{\rm G}\over (30\ {\rm yr})^{-1}}\right]
\left({M_{\rm mix}\over 3\times 10^4\,M_\odot}\right)
\left({10^{10}\,M_\odot\over M_{\rm gas}}\right).
\ee
The present Galactic rate of $\sim (30\ {\rm yr})^{-1}$ for core-collapse
supernovae \cite{ca99} corresponds to 
$f_{\rm mix}^{\rm SN}\sim (10^7\ {\rm yr})^{-1}$ while that of
$\sim (10^5\ {\rm yr})^{-1}$ for neutron star mergers (e.g., \cite{ph91})
corresponds to 
$f_{\rm mix}^{\rm NSM}\sim (3\times 10^{10}\ {\rm yr})^{-1}$.
As $f_H>f_L\gg f_{\rm mix}^{\rm NSM}$,
it seems that neutron star mergers cannot be associated with either 
the $H$ or $L$ events. This conclusion is unaffected if the upper limit
of $\sim (10^4\ {\rm yr})^{-1}$ \cite{acw99} instead of the nominal value
of $\sim (10^5\ {\rm yr})^{-1}$ is used for the present Galactic rate of
neutron star mergers. By contrast, $f_{\rm mix}^{\rm SN}$ is comparable
to $f_H$ and $f_L$ within a factor of $\sim 10$, 
which suggests that the $H$ and $L$ events may be associated 
with core-collapse supernovae \cite{wa96,qi00}. It is important that
the above argument be substantiated by more detailed studies on the mixing 
of the ejecta from core-collapse supernovae and neutron star mergers with 
the ISM and on the occurrences of these events over the 
Galactic history.
Independent of the association with specific astrophysical sites, 
the characteristics of the $H$ and $L$ events are
further tested and elucidated by observations of abundances
in metal-poor stars.

\subsection{Observations of abundances in metal-poor stars}
The effects of diverse sources for the $r$-process would be most prominent
in the early Galaxy where not many $r$-process events had contributed to
the abundances in a reference ISM. The chemical evolution of the early
Galaxy is reflected by the abundances in metal-poor stars that reside in
the Galactic halo. Observations of abundances in a large number of 
metal-poor stars as well as detailed studies covering many elements in
individual stars have been carried out by a number of groups (e.g., 
\cite{mc95}--\cite{car02}).
These observations provide strong support for the diversity of $r$-process
sources and give important information on the $r$-patterns produced
by these sources. 

Except for the trivial case where an element has a single $r$-process 
isotope and some special cases \cite{ma95}--\cite{la02}
where the hyperfine structure of the
atomic spectra allows the extraction of the isotopic composition
(see discussion of Ba in \S\ref{fis}), stellar
observations can only give the total abundance of all the isotopes
of an element. The observed abundance of an element E in a star is
usually given in terms of
\be
\log\epsilon({\rm E})=\log({\rm E/H})+12,
\ee
where (E/H) is the abundance ratio of the element E relative to hydrogen
in the star. Hydrogen is an excellent reference element as its overall
abundance has changed little over the history of the universe. The
``metallicity'' of a star is measured by
\be
[{\rm Fe/H}]=\log\epsilon({\rm Fe})-\log\epsilon_\odot({\rm Fe})
=\log({\rm Fe/H})-\log({\rm Fe/H})_\odot,
\ee
where the subscript ``$\odot$'' denotes quantities for the sun.
There are two possible sources for Fe: SNe II and Ia. However, as SNe Ia
are associated with low-mass progenitors that evolve on timescales of
$\sim 10^9$ yr, only SNe II with short-lived massive progenitors 
contributed Fe during the first $\sim 10^9$ yr subsequent to the onset of
Galactic chemical evolution ($t=0$). Over a period of $\sim 10^{10}$ yr
prior to the formation of the solar system, SNe II contributed $\sim 1/3$
of the solar Fe abundance (e.g., \cite{ti95}). This gives
\be
\left({{\rm Fe}\over{\rm H}}\right)\sim {1\over 3}
\left({{\rm Fe}\over{\rm H}}\right)_\odot
\left({t\over 10^{10}\ {\rm yr}}\right)
\ee
for $t\lesssim 10^9$ yr. Thus, for stars with [Fe/H]~$\lesssim -1.5$
corresponding to $t\lesssim 10^9$ yr, their Fe abundances are dominated
by SN II contributions. Stars with 
$-3\lesssim [{\rm Fe/H}]< -2.5$ are 
referred to as ultra-metal-poor (UMP) stars and those with 
$-2.5\lesssim [{\rm Fe/H}]\lesssim -1.5$ as metal-poor (MP) stars.

The observational data of interest here concern the elements Sr and above.
In general, both the $r$-process and the main $s$-process can produce
these elements. However, the main $s$-process contributions dominantly
come from low-mass AGB stars that evolve on timescales of $\sim 10^9$ yr.
So the abundances of the elements Sr and above in UMP and MP stars 
represent the $r$-process contributions as pointed out by Truran 
\cite{tr81}. The occurrence of $r$-process events in the early
Galaxy ($t\lesssim 10^9$ yr) suggests that these events are associated 
with objects such as SNe II, which have short-lived massive progenitors.

\subsubsection{nonsolar $r$-patterns in UMP stars\label{nonsol}}
The observed abundances of a large number of elements in the UMP star
CS 22892--052 \cite{sn96,sn00} are shown in Fig. 7. The solar $r$-pattern
\cite{ar99} translated to pass through the Eu data is also shown for
comparison. It can be seen that the data on the heavy $r$-process elements
from Ba to Ir are in excellent agreement with 
the translated solar $r$-pattern.
However, there are large differences between the data and this pattern
in the region of the light $r$-process elements below Ba.
It is important to examine how the uncertainties in the solar $r$-process
abundances may affect the above comparison.
The fraction of the solar abundance contributed by the
$s$-process is very small ($\beta_{\odot,s}\lesssim 0.3$) for the 
following groups of elements: (1) Ru, Rh, and Ag; (2) from Sm to Yb;
and (3) Os and Ir \cite{ar99} (see Table 1 in \S\ref{rhl}). 
The solar $r$-process abundances of these
elements should be rather reliable. Thus, in comparing the data for
CS 22892--052 and the
translated solar $r$-pattern, the agreement for the elements above Ba
in groups (2) and (3) and the disagreement 
for Rh and Ag below Ba in group (1) 
are robust. Regardless of the possible uncertainties in the solar 
$r$-process abundances of the other elements (see discussion of Sr, Y,
Zr, Nb, and Ba in \S\ref{rhl}), the observed $r$-pattern in
CS 22892--052 is clearly different from the overall solar $r$-pattern.
Similar difference is also observed for another UMP star CS 31082--001
\cite{hi02}. These observations provide strong support for the 
diversity of $r$-process sources as concluded from the meteoritic data on 
$^{182}$Hf and $^{129}$I. The observed deficiency especially at the light 
$r$-process elements Rh and Ag relative to the solar $r$-pattern that is
translated to pass through the Eu data reflects the characteristics
of the $H$ events that produce mainly the heavy $r$-process nuclei.
By inference, a mixture of the $H$ events and the $L$ events that produce 
dominantly the light $r$-process nuclei is then required to account for 
the overall solar $r$-pattern.

\begin{figure}[tb]
\begin{minipage}{2 cm}
\makebox[1cm]{}
\end{minipage}
\begin{minipage}{8 cm}
\epsfig{file=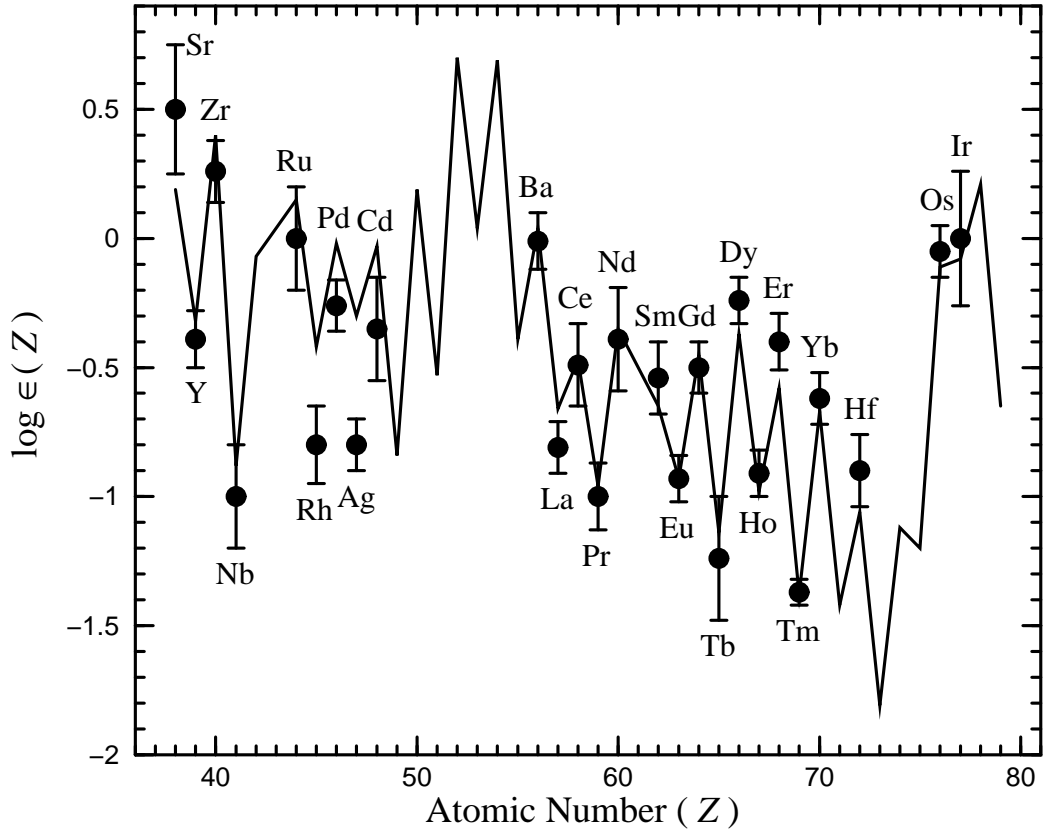,angle=270,scale=0.7}
\end{minipage}
\begin{center}
\begin{minipage}{16.5 cm}
\caption{The observed abundances in CS 22892--052 (filled circles with error 
bars: \cite{sn96,sn00}) compared with the solar $r$-pattern (solid curve: 
\cite{ar99}) that is translated to pass through the Eu data. 
The data on the heavy 
$r$-process elements from Ba to Ir are in excellent agreement with
the translated solar 
$r$-pattern. However, the data on the light $r$-process 
elements Rh and Ag clearly fall below this pattern. See text for details.
See also \S\ref{rhl} for discussion of corrections to the solar
$r$-process abundances of Sr, Y, Zr, Nb, and Ba.}
\end{minipage}
\end{center}
\end{figure}

\subsubsection{yields of $H$ and $L$ events and three-component
model for abundances in UMP and MP stars\label{rhl}}
The differences in the observed $r$-process abundances between the UMP 
stars HD 115444 \cite{we00}, CS 31082--001 \cite{hi02}, and CS 22892--052
\cite{sn96,sn00} are shown as $\Delta\log\epsilon(Z)=
\log\epsilon_{\rm HD115444}(Z)-\log\epsilon_{\rm CS22892}(Z)$ or
$\log\epsilon_{\rm CS31082}(Z)-\log\epsilon_{\rm CS22892}(Z)$
in Fig. 8. It can be seen that within the
observational uncertainties, these differences represent a uniform shift
in $\log\epsilon(Z)$ 
of $\approx -0.6$ dex for HD 115444 and of $\approx 0.21$ dex for
CS 31082--001, both relative to CS 22892--052. This means that these three
UMP stars have essentially the same $r$-pattern. This pattern is deficient
in the light $r$-process elements such as Rh and Ag 
relative to the solar $r$-pattern translated to pass through the Eu data 
(an established result for CS 22892--052, a confirmed prediction for
CS 31082--001 \cite{hi02,qw01a}, and still a prediction for HD 115444),
and therefore, 
is characteristic of the $H$ events (see \S\ref{nonsol}). The $H$ events 
then have an approximately fixed yield pattern, 
which can be taken from the data 
for e.g., CS 22892--052. As the $H$ yield pattern coincides with the solar
$r$-pattern and the $r$-patterns in a number of MP stars
\cite{sn98,jo01} in the region of the heavy $r$-process elements, it is
reasonable to assume that these elements are produced exclusively by the
$H$ events. In this case, the occurrence of these events can be 
represented by the abundance of a typical heavy $r$-process
element such as Eu.

\begin{figure}[tb]
\begin{minipage}{2 cm}
\makebox[1cm]{}
\end{minipage}
\begin{minipage}{8 cm}
\epsfig{file=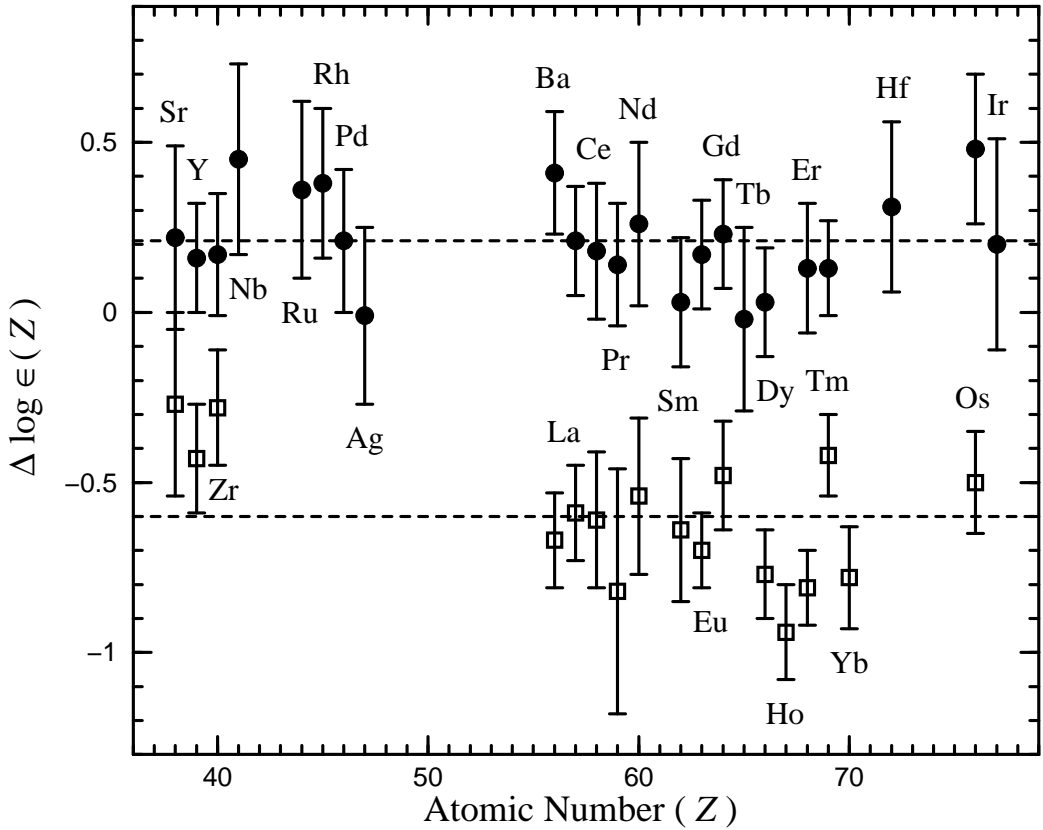,angle=270,scale=0.7}
\end{minipage}
\begin{center}
\begin{minipage}{16.5 cm}
\caption{Differences in the observed $r$-process abundances between
HD 115444 \cite{we00} and CS 22892--052 \cite{sn96,sn00}
(open squares with error 
bars) and those between CS 31082--001 \cite{hi02} and CS 22892--052 (filled 
circles with error bars). The data are consistent with a uniform shift 
in $\log\epsilon(Z)$ of $\approx -0.6$ dex (lower dashed line) for 
HD 115444 and of $\approx 0.21$
dex (upper dashed line) for CS 31082--001, both relative to CS 22892--052.
This suggests an approximately fixed yield pattern for the $H$ events.
The upward offset from the lower dashed line for the 
elements Sr, Y, and Zr in HD 115444 indicates that this star may 
have received large contributions to these elements from the $P$ 
inventory [to be discussed in the text after Eq. (\ref{epsfel})]
in addition to the $H$ events. By contrast, the abundances
of Sr, Y, and Zr in CS 31082--001 and CS 22892--052 are dominated by
the contributions from the $H$ events.}
\end{minipage}
\end{center}
\end{figure}

Over the period of $T_{\rm UP}\sim 10^{10}$ yr prior to the formation of
the solar system, a total number $n_H^\odot=f_HT_{\rm UP}\sim 10^3$ of $H$
events contributed to the solar $r$-process abundance of Eu
for $f_H\sim (10^7\ {\rm yr})^{-1}$. This gives
(Eu/H)$_{\odot,r}=n_H^\odot({\rm Eu/H})_H$, or
\be
\log\epsilon_H({\rm Eu})=\log\epsilon_{\odot,r}({\rm Eu})-\log n_H^\odot
\sim -2.5,
\label{epseuh}
\ee
where $\log\epsilon_{\odot,r}({\rm Eu})=0.52$ corresponds to the solar 
$r$-process abundance of Eu \cite{ar99} and $\log\epsilon_H({\rm Eu})$
represents the Eu abundance resulting from a single $H$ event. The data
on Eu over the wide range of $-3\lesssim [{\rm Fe/H}]\lesssim -1$ are
shown in Fig. 9. The lowest observed Eu abundances are consistent with
the enrichment level of $\log\epsilon_H({\rm Eu})\sim -2.5$ for a single
$H$ event. This consistency further establishes that the frequency of the
$H$ events is $f_H\sim (10^7\ {\rm yr})^{-1}$, which is close to the 
frequency for replenishing newly-synthesized material in the ISM by
core-collapse supernovae (see \S\ref{met}). If 
$f_H$ were $\sim (10^{10}\ {\rm yr})^{-1}$, which is close to the 
frequency for replenishing newly-synthesized material in the ISM by 
neutron star mergers (see \S\ref{met}), $n_H^\odot$ would be $\sim 1$.
The corresponding enrichment level of $\log\epsilon_H({\rm Eu})\sim 0.52$
for a single $H$ event [see Eq. (\ref{epseuh})] would be in clear conflict 
with the Eu data shown in Fig. 9. 
This again suggests that the $H$ events cannot 
be associated with neutron star mergers \cite{qi00}. A similar argument 
against the association of the $L$ events with neutron star mergers 
\cite{qi00} can be made by using the data on the light $r$-process 
element Ag \cite{cr98}. It is important that these arguments be verified
by more sophisticated studies on the inhomogeneous chemical evolution
of the early Galaxy.

\begin{figure}[tb]
\begin{minipage}{2 cm}
\makebox[1cm]{}
\end{minipage}
\begin{minipage}{8 cm}
\epsfig{file=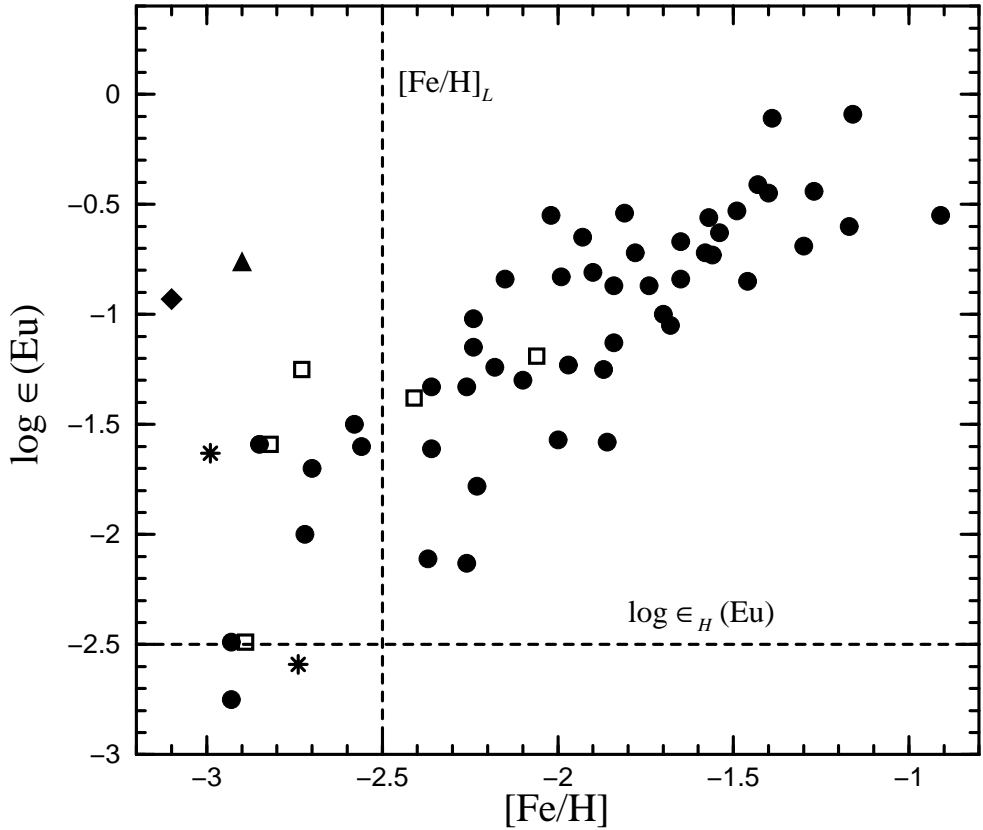,angle=270,scale=0.7}
\end{minipage}
\begin{center}
\begin{minipage}{16.5 cm}
\caption{Data on Eu over the wide range of 
$-3\lesssim [{\rm Fe/H}]\lesssim -1$ (filled circles: \cite{bur00};
open squares: \cite{mc95}; asterisks: \cite{we00}; filled diamond: 
\cite{sn00}; filled triangle: \cite{hi02}). There is a large dispersion
of $\sim 2$ dex in $\log\epsilon({\rm Eu})$ at 
$-3\lesssim [{\rm Fe/H}]< -2.5$, which suggests that
the $H$ events responsible for Eu enrichment cannot produce any significant 
amount of Fe. The Eu abundance resulting from a 
single $H$ event (horizontal dashed line) is consistent with
the lowest observed Eu abundances. The vertical dashed line indicates
the Fe abundance resulting from a single $L$ event. See text for details.}
\end{minipage}
\end{center}
\end{figure}

Figure 9 shows that there is a large dispersion of $\sim 2$ dex in 
$\log\epsilon({\rm Eu})$ over the narrow range of 
$-3\lesssim [{\rm Fe/H}]< -2.5$. In particular, CS 31082--001 and 
CS 22892--052 with the highest Eu abundances among the UMP stars shown
in this figure have very low values of [Fe/H]~$=-2.9$ and $-3.1$,
respectively. This suggests that the $H$ events responsible for Eu 
enrichment cannot produce any significant amount of Fe \cite{wa00}.
No Fe is expected to be produced in neutron star mergers. However, as
argued above, the rarity of these events appears to result in
conflicts with both the meteoritic data on $^{182}$Hf and observations 
of abundances
in metal-poor stars if they are associated with the $H$ events. While
the possibility of neutron star mergers being an $r$-process source
merits further studies, the discussion below assumes that the $H$ and
$L$ events are associated with core-collapse supernovae, which include
SNe II from the collapse of O-Ne-Mg and Fe cores as well as the
silent supernovae from
accretion-induced collapse of white dwarfs. A number of SNe II including 
SN 1987A are known to produce Fe (see e.g., Table 1 of \cite{so02}). 
In fact, $\sim 1/3$ of the solar
Fe abundance is attributed to SNe II (e.g., \cite{ti95}). 
As no Fe can be produced by the $H$ events, this Fe must be
assigned to a total number $n_L^\odot=f_LT_{\rm UP}\sim 10^2$ of $L$ 
events that occurred with a frequency of $f_L\sim (10^8\ {\rm yr})^{-1}$
over the period of $T_{\rm UP}\sim 10^{10}$ yr prior
to the formation of the solar system. This gives 
$(1/3){\rm (Fe/H)}_\odot\sim n_L^\odot{\rm (Fe/H)}_L$, or
\be
{\rm [Fe/H]}_L\sim -\log(3n_L^\odot)\sim -2.5,
\label{epsfel}
\ee
where [Fe/H]$_L$ represents the Fe abundance resulting from a single $L$
event.

The UMP stars have $-3\lesssim [{\rm Fe/H}]< -2.5$, and therefore, cannot
have received any contributions from the $L$ events. The Fe in these stars
is attributed to the very massive ($\gtrsim 100\,M_\odot$) stars that
dominated the chemical evolution of the universe until [Fe/H]~$\sim -3$
was reached \cite{wa00}. At this critical metallicity, very massive stars
could not be formed any more and major formation of regular 
($\sim 1$--$60\,M_\odot$) stars took over. In addition to Fe, other 
elements were also produced by the very massive stars.
The inventory of such elements associated with the production of
[Fe/H]~$\sim -3$ is referred to as the prompt ($P$)
inventory. The abundance of an element E in a UMP star represents the sum
of the $P$ inventory and the contributions from the $H$ events:
\be
{\rm (E/H)}={\rm (E/H)}_P+{\rm (E/Eu)}_H{\rm (Eu/H)},
\label{eump}
\ee
where (E/H)$_P$ is the $P$ inventory of E and (E/Eu)$_H$ is the relative
$H$ yield of E to Eu. The parameter (E/Eu)$_H$ is assumed to be the same
for all $H$ events (see Fig. 8). 
Then the parameters (E/H)$_P$ and (E/Eu)$_H$ can be 
obtained from the data for two UMP stars with essentially the same [Fe/H]
but very different Eu abundances (e.g., CS 22892--052 and HD 115444).
It was found that there is a significant $P$ inventory for the light
$r$-process elements Sr, Y, and Zr \cite{qw01}. The parameters (E/H)$_P$ 
and (E/Eu)$_H$ for these elements are
given in terms of $\log\epsilon_P({\rm E})$ and $\log\epsilon_H({\rm E})$
in Table 1. The parameter (E/Eu)$_H$ can be calculated as
\be
\left({{\rm E}\over{\rm Eu}}\right)_H=
10^{\log\epsilon_H({\rm E})-\log\epsilon_H({\rm Eu})},
\label{heeu}
\ee
where $\log\epsilon_H({\rm E})$ represents the abundance of E resulting
from a single $H$ event (the $H$ yield of E)
and $\log\epsilon_H({\rm Eu})=-2.48$ corresponds
to assigning the solar $r$-process abundance of Eu to $n_H^\odot=10^3$
$H$ events [see Eq. (\ref{epseuh})]. By assuming that the $P$ inventory
is negligible for the elements above Zr, the $\log\epsilon_H$ values
for the elements from Nb to Cd can be calculated from the data for
CS 22892--052. As the heavy $r$-process elements Ba and above follow
the solar $r$-pattern, the $\log\epsilon_H$ values for these elements
can be calculated from the data for CS 22892--052 or from the solar
$r$-process abundances as in the case of Eu [see Eq. (\ref{epseuh})]. 
The $\log\epsilon_H$ values for the elements above Zr are also given 
in Table 1. By applying Eq. (\ref{eump}) to the data for UMP stars,
it was further found that the $H$ events cannot produce any significant 
amount of the elements from Na to Ni, which include not only the Fe 
group but also the so-called ``$\alpha$-elements'' such as
Mg, Si, Ca, and Ti \cite{qw02}.

\begin{table}
\begin{center}
\begin{minipage}[t]{16.5 cm}
\caption{Parameters of the three-component model for abundances in
UMP and MP stars (see Table 3 of \cite{qw01}). The $P$ inventory in
col. (2) and the $H$ yields in col. (3) are calculated from the data
on HD 115444 \cite{we00} and CS 22892--052 \cite{sn00} for Sr, Y, and Zr.
The $P$ inventory for the elements above Zr is assumed to be 
negligible and set to $-\infty$. The $H$ yields of Nb to Cd are calculated 
from the data for CS 22892--052 and those of Ba to Au from the standard 
solar $r$-process abundances \cite{ar99} in col. (5) [corresponding to the 
solar $s$-process fractions in col. (6)]. The $L$ yields in col. (4) are 
calculated from cols. (2), (3), and (5) for Sr to Cd and set to $-\infty$ 
for Ba and above. The corrected values for Sr, Y, Zr, and Ba are given 
in cols. (7)--(9). These columns are omitted for La to Au, for which
elements no corrections are recommended.}
\end{minipage}
\begin{tabular}{crrrrrrrr}
&&&&&&&&\\[-2mm]\hline
&&&&&&&&\\[-2mm]
E&$\log\epsilon_P({\rm E})$&$\log\epsilon_H({\rm E})$&
$\log\epsilon_L({\rm E})$&$\log\epsilon_{\odot,r}({\rm E})$&
$\beta_{\odot,s}({\rm E})$&$\log\epsilon_L^{\rm corr}({\rm E})$&
$\log\epsilon_{\odot,r}^{\rm corr}({\rm E})$&
$\beta_{\odot,s}^{\rm corr}({\rm E})$\\
(1)&(2)&(3)&(4)&(5)&(6)&(7)&(8)&(9)\\
&&&&&&&&\\[-2mm]
\hline
&&&&&&&&\\[-2mm]
Fe&4.51$^{\rm a}$&$-\infty$&5.03$^{\rm a}$&---&---&---&---&---\\[2mm]
Sr&0.13&$-1.30$&$-\infty^{\rm b}$&1.64&0.95&0.35&2.44&0.69\\[2mm]
Y&$-1.05$&$-2.05$&$-1.36$&1.12&0.92&$-0.42$&1.67&0.72\\[2mm]
Zr&$-0.13$&$-1.53$&$-0.40$&1.85&0.83&0.26&2.32&0.49\\[2mm]
Nb&$-\infty$&$-2.56$&$-2.06$&0.56&0.85&
---$^{\rm c}$&---$^{\rm c}$&---$^{\rm c}$\\[2mm]
Ru&$-\infty$&$-1.56$&$-0.92$&1.60&0.32&---&---&---\\[2mm]
Rh&$-\infty$&$-2.36$&$-1.20$&1.03&0.14&---&---&---\\[2mm]
Pd&$-\infty$&$-1.82$&$-0.95$&1.42&0.46&---&---&---\\[2mm]
Ag&$-\infty$&$-2.36$&$-1.02$&1.14&0.20&---&---&---\\[2mm]
Cd&$-\infty$&$-1.91$&$-0.85$&1.42&0.52&---&---&---\\[2mm]
Ba&$-\infty$&$-1.52^{\rm d}$&$-\infty$&1.48&0.81&$-0.47$&1.79&0.62\\[2mm]
&&&&&&&&\\[-2mm]\hline
\end{tabular}
\begin{minipage}[t]{16.5 cm}
\vskip 0.5cm
\noindent
$^{\rm a}$ With $\log\epsilon_\odot({\rm Fe})=7.51$ \cite{an89},
these values correspond to [Fe/H]$_P=-3$ and [Fe/H]$_L=-2.48$.\\
$^{\rm b}$ The standard solar $r$-process abundance of Sr is saturated by
the contributions from $n_H^\odot=10^3$ $H$ events, and therefore,
does not allow any 
contributions to Sr from the $L$ events.\\
$^{\rm c}$ The data for BD +17$^\circ$3248 \cite{co02} suggest
$\log\epsilon_L^{\rm corr}({\rm Nb})\approx -0.88$ to $-0.66$, which 
corresponds to $\log\epsilon_{\odot,r}^{\rm corr}({\rm Nb})\approx 1.20$ 
to $1.39$ and $\beta_{\odot,s}({\rm Nb})\approx 0.37$ to 0.02. This
corrected $L$ yield of Nb should be tested by future data for MP stars.\\
$^{\rm d}$ This $H$ yield of Ba, 
which is calculated by attributing its standard 
solar $r$-process abundance to $n_H^\odot=10^3$ $H$ events, 
essentially coincides with the value of 
$\log\epsilon_H({\rm Ba})\approx -1.57$ obtained from the data for 
CS 22892--052. The use of the latter value is recommended.
\end{minipage}
\end{center}
\end{table}

\begin{table}
\begin{center}
\begin{minipage}[t]{16.5 cm}
\centerline{Table 1 (continued)}
\end{minipage}
\begin{tabular}{crrrrr}
&&&&&\\[-2mm]\hline
&&&&&\\[-2mm]
E&$\log\epsilon_P({\rm E})$&$\log\epsilon_H({\rm E})$&
$\log\epsilon_L({\rm E})$&$\log\epsilon_{\odot,r}({\rm E})$&
$\beta_{\odot,s}({\rm E})$\\
(1)&(2)&(3)&(4)&(5)&(6)\\
&&&&&\\[-2mm]
\hline
&&&&&\\[-2mm]
La&$-\infty$&$-2.22$&$-\infty$&0.78&0.62\\[2mm]
Ce&$-\infty$&$-2.02$&$-\infty$&0.98&0.77\\[2mm]
Pr&$-\infty$&$-2.51$&$-\infty$&0.49&0.49\\[2mm]
Nd&$-\infty$&$-1.90$&$-\infty$&1.10&0.56\\[2mm]
Sm&$-\infty$&$-2.20$&$-\infty$&0.80&0.29\\[2mm]
Eu&$-\infty$&$-2.48$&$-\infty$&0.52&0.06\\[2mm]
Gd&$-\infty$&$-2.00$&$-\infty$&1.00&0.15\\[2mm]
Tb&$-\infty$&$-2.70$&$-\infty$&0.30&0.07\\[2mm]
Dy&$-\infty$&$-1.92$&$-\infty$&1.08&0.15\\[2mm]
Ho&$-\infty$&$-2.53$&$-\infty$&0.47&0.08\\[2mm]
Er&$-\infty$&$-2.13$&$-\infty$&0.87&0.17\\[2mm]
Tm&$-\infty$&$-2.93$&$-\infty$&0.07&0.13\\[2mm]
Yb&$-\infty$&$-2.22$&$-\infty$&0.78&0.33\\[2mm]
Lu&$-\infty$&$-2.98$&$-\infty$&0.02&0.20\\[2mm]
Hf&$-\infty$&$-2.61$&$-\infty$&0.39&0.56\\[2mm]
Ta&$-\infty$&$-3.36$&$-\infty$&$-0.36$&0.41\\[2mm]
W&$-\infty$&$-2.68$&$-\infty$&0.32&0.56\\[2mm]
Re&$-\infty$&$-2.75$&$-\infty$&0.25&0.09\\[2mm]
Os&$-\infty$&$-1.66$&$-\infty$&1.34&0.09\\[2mm]
Ir&$-\infty$&$-1.63$&$-\infty$&1.37&0.01\\[2mm]
Pt&$-\infty$&$-1.34$&$-\infty$&1.66&0.05\\[2mm]
Au&$-\infty$&$-2.20$&$-\infty$&0.80&0.06\\[2mm]
&&&&&\\[-2mm]\hline
\end{tabular}
\end{center}
\end{table}

The Fe at $-2.5\lesssim [{\rm Fe/H}]\lesssim -1.5$ represents the
contributions from the $L$ events added to the $P$ inventory. The
abundance of an element E in an MP star in this metallicity range
can be written as
\be
{\rm (E/H)}={\rm (E/H)}_P+{\rm (E/Eu)}_H{\rm (Eu/H)}+
{\rm (E/Fe)}_L[{\rm (Fe/H)}-{\rm (Fe/H)}_P],
\label{emp}
\ee
where (E/Fe)$_L$ is the relative $L$ yield of E to Fe and is assumed
to be fixed for all $L$ events. With the
parameters (E/H)$_P$ and (E/Eu)$_H$ given in terms of
$\log\epsilon_P({\rm E})$ and $\log\epsilon_H({\rm E})$ in
Table 1, the parameter
(E/Fe)$_L$ can be obtained from the solar $r$-process abundance of E:
\be
{\rm (E/H)}_{\odot,r}={\rm (E/H)}_P+{\rm (E/Eu)}_H{\rm (Eu/H)}_{\odot,r}
+{\rm (E/Fe)}_L(1/3){\rm (Fe/H)}_\odot,
\label{yl}
\ee
where 1/3 of the solar Fe abundance is taken to be contributed by the
$L$ events. The parameter (E/Fe)$_L$ 
calculated in this way for the light $r$-process elements from Sr to
Cd are given in terms of $\log\epsilon_L({\rm E})$ in Table 1.
The parameter (E/Fe)$_L$ can be calculated as
\be
\left({{\rm E}\over{\rm Fe}}\right)_L=
10^{\log\epsilon_L({\rm E})-\log\epsilon_L({\rm Fe})},
\label{lefe}
\ee
where $\log\epsilon_L({\rm E})$ represents the abundance of E resulting
from a single $L$ event (the $L$ yield of E)
and $\log\epsilon_L({\rm Fe})=5.03$ 
([Fe/H]$_L=-2.48$) corresponds
to assigning 1/3 of the solar Fe abundance to $n_L^\odot=10^2$
$L$ events [see Eq. (\ref{epsfel})].
With the parameters $\log\epsilon_P({\rm E})$, $\log\epsilon_H({\rm E})$,
and $\log\epsilon_L({\rm E})$ given in Table 1,
the abundance of E in any UMP or MP star can be calculated from Eqs.
(\ref{eump})--(\ref{emp}) and (\ref{lefe}) 
by using only the observed abundances of Eu 
and Fe in the star. This is the three-component ($P$, $H$, and $L$) model 
for abundances in UMP and MP stars as developed by Qian and Wasserburg
\cite{qw01}.

By comparing the data and the model predictions, large discrepancies
were found for the elements Sr, Y, Zr, and Ba at
$-2.5\lesssim [{\rm Fe/H}]\lesssim -1.5$. One possible explanation is that 
significant $s$-process contributions to these elements already occurred
at metallicities as low as [Fe/H]~$\sim -2.5$ (e.g., \cite{bur00}) 
but were not taken into account by the three-component model. 
However, it is also possible that the
contributions from the $L$ events to the elements Sr, Y, and Zr were
greatly underestimated by the model and that the $L$ 
events may produce a large amount of Ba. Good agreement 
between the data and the model can be obtained when the solar 
$r$-process abundances of Sr, Y, Zr, and Ba are increased from their
standard values \cite{ar99} by factors of $\sim 2$--6 and the $L$ yields
of these elements are correspondingly increased \cite{qw01}.
The solar abundances of Sr, 
Y, Zr, and Ba are dominated by the $s$-process contributions 
($\beta_{\odot,s}\gtrsim 0.8$ \cite{ar99}; see Table 1). 
Thus, the solar $r$-process 
abundances of these elements obtained by subtracting the $s$-process
contributions are subject to possibly large uncertainties 
(see \S\ref{sintro}). It remains to be seen
if future $s$-process studies can justify the corrected solar $r$-process 
abundances obtained by comparing the data and the three-component model
for abundances in UMP and MP stars. The
corrected parameters are given in Table 1. 
The corrected $L$ yield of Ba is $\approx 10$ times
larger than the $H$ yield of Ba. However,
the heavy $r$-process elements above Ba are still produced 
exclusively by the $H$ events. This suggests that the $L$ events
can only produce the elements up to and including Ba.

As an example to test the three-component model (with corrected $L$ yields
of Sr, Y, Zr, and Ba),
the abundances calculated for BD +17$^\circ$3248 by using the observed
values of $\log\epsilon({\rm Eu})=-0.67$ and 
$\log\epsilon({\rm Fe})=5.43$ are compared with the data \cite{co02} in
Fig. 10. It can be seen that the agreement between the model and the data 
is quite good except for the element Nb. However, as in the case of the 
elements Sr, Y, Zr, and Ba, the solar abundance of Nb is dominated by
the $s$-process contributions 
($\beta_{\odot,s}=0.85$ \cite{ar99}; see Table 1).
Thus, the $L$ yield of Nb derived from Eq. (\ref{yl}) 
by using its solar $r$-process abundance may be quite uncertain. 
The data for BD +17$^\circ$3248 suggest a corrected $L$ yield of 
$\log\epsilon_L^{\rm corr}({\rm Nb})\approx -0.88$ to $-0.66$, which
corresponds to a rather small value of
$\beta_{\odot,s}^{\rm corr}({\rm Nb})\approx 0.37$ to 0.02. These corrected
values should be tested by future data for MP stars and detailed
$s$-process models.

\begin{figure}[tb]
\begin{minipage}{2 cm}
\makebox[1cm]{}
\end{minipage}
\begin{minipage}{8 cm}
\epsfig{file=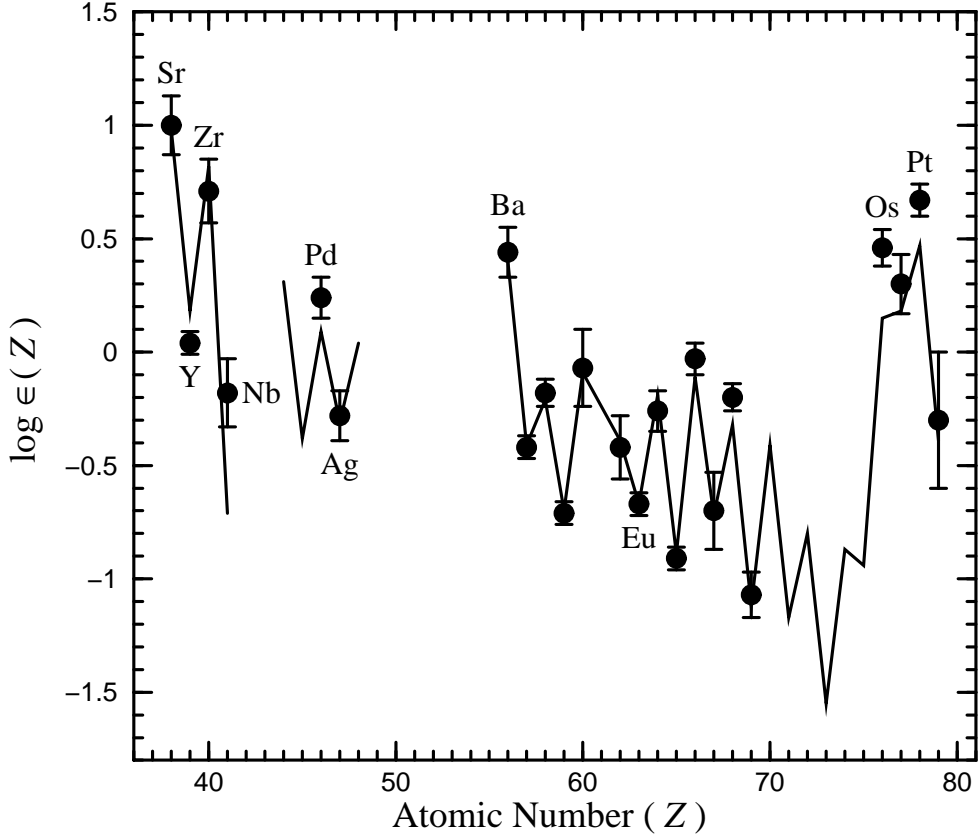,angle=270,scale=0.7}
\end{minipage}
\begin{center}
\begin{minipage}{16.5 cm}
\caption{The observed abundances in BD +17$^\circ$3248 (filled circles with
error bars: \cite{co02}) compared with the predictions from the 
three-component model (solid curves). The model \cite{qw01} uses only the 
data on Eu and Fe to calculate the abundances of all the other elements. The 
agreement between the model and the data is quite good except for the element 
Nb. The $L$ yield of Nb may have been underestimated by the model due to the
uncertainty in the solar $r$-process abundance of Nb.
See text for details.}
\end{minipage}
\end{center}
\end{figure}

\subsubsection{variations in the $H$ yield pattern and nucleochronology
\label{chron}}
The discussion of stellar observations has focused on the stable elements
so far. Observations of the long-lived radioactive elements Th and U in
metal-poor stars are considered below in connection with nucleochronology.
The dominant radioactive nuclei in stars with an age of 
$t_{\rm star}\gtrsim 10^{10}$ yr
are $^{232}$Th (with a lifetime of $\tau_{232}=2.03\times 10^{10}$ yr) and
$^{238}$U ($\tau_{238}=6.45\times 10^9$ yr). The period over which the 
$r$-process events could contribute to the initial abundances of 
these two nuclei in such old stars is limited by the age of the 
universe ($\approx 1.5\times 10^{10}$ yr) to shorter than the lifetime 
of either nucleus. So the initial abundance ratio 
($^{238}$U/$^{232}$Th)$_0$ in these stars
essentially reflects the relative yield of
$^{238}$U to $^{232}$Th for the $r$-process events. As these two nuclei
are close in mass number, their relative yield $Y_{238}/Y_{232}$ may
be rather constant among individual events. Then the present abundance
ratio ($^{238}$U/$^{232}$Th)$_{\rm star}$ in an old star is
\be
\left({^{238}{\rm U}\over ^{232}{\rm Th}}\right)_{\rm star} =
\left({^{238}{\rm U}\over ^{232}{\rm Th}}\right)_0
\exp\left[-\left({t_{\rm star}\over\tau_{238}}-
{t_{\rm star}\over\tau_{232}}\right)\right]
\approx\left({Y_{238}\over Y_{232}}\right)
\exp\left[-\left({t_{\rm star}\over\tau_{238}}-
{t_{\rm star}\over\tau_{232}}\right)\right],
\ee
which gives
\be
t_{\rm star}\approx (21.8\times 10^9\ {\rm yr})
[\log(Y_{238}/Y_{232})-\log({\rm U/Th})_{\rm star}],
\label{uth}
\ee
where the present abundance
ratio ($^{238}$U/$^{232}$Th)$_{\rm star}$ in the star has been set to 
be the same as the elemental abundance ratio (U/Th)$_{\rm star}$. 
In principle, $Y_{238}/Y_{232}$
can be calculated from theory and various estimates give 
$Y_{238}/Y_{232}\approx 0.53\pm 0.23$ (e.g., \cite{go01}). 
If the abundances of Th and U are measured
in the star, then its age can be estimated from Eq. (\ref{uth}). So far,
the abundances of both Th and U have been observed only in one
UMP star, CS 31082--001 \cite{hi02,cay01}, and one MP
star, BD +17$^\circ$3248 \cite{co02}. 
Age estimates based on Eq. (\ref{uth}) are rather unreliable for 
BD +17$^\circ$3248 mainly due to the large observational uncertainty of 
$\sim 0.3$ dex in $\log\epsilon({\rm U})$ for this star. By contrast,
an age of $(13.5\pm 5.0)\times 10^9$ yr can be estimated from
$\log({\rm U/Th})=-0.94\pm 0.11$ for CS 31082--001 \cite{hi02} 
(see also \cite{cay01}), where the error in the age is dominated by the
uncertainty in the theoretical estimate of $Y_{238}/Y_{232}$.

If the heavy $r$-process elements including the long-lived chronometers
are exclusively produced by the $H$ events and the $H$ yield pattern is
fixed, then nucleochronology can be applied to stars for which the
abundances of Th and a stable heavy $r$-process element such as Eu are
known (e.g., \cite{we00,jo01,co02,cow97,cow99}). The present 
abundance ratio (Th/Eu)$_{\rm star}$ in a star is related to its age by
\be
\left({{\rm Th}\over {\rm Eu}}\right)_{\rm star} =
\left({{\rm Th}\over {\rm Eu}}\right)_0
\exp\left(-{t_{\rm star}\over\tau_{232}}\right),
\label{theu}
\ee
which gives
\be
t_{\rm star}= (46.7\times 10^9\ {\rm yr})
[\log({\rm Th/Eu})_0-\log({\rm Th/Eu})_{\rm star}].
\label{ttheu}
\ee
The initial abundance ratio (Th/Eu)$_0$ in Eqs. (\ref{theu}) and 
(\ref{ttheu}) is essentially the relative
$H$ yield of Th to Eu, (Th/Eu)$_H$, and can be estimated from the 
$r$-process abundance ratio (Th/Eu)$_{{\rm ESS},r}$ 
in the early solar system.
Due to the long lifetime of $^{232}$Th, the replenishment of this nucleus
in the ISM by the $H$ events can be considered as continuous and the 
change in (Th/H) is governed by
\be
{d\over dt}\left({{\rm Th}\over {\rm H}}\right)=
f_H\left({{\rm Th}\over {\rm H}}\right)_H-
{1\over\tau_{232}}\left({{\rm Th}\over {\rm H}}\right),
\ee
which can be solved to give
\be
\left({{\rm Th}\over {\rm H}}\right)_{\rm ESS}=
\left({{\rm Th}\over {\rm H}}\right)_Hf_H\tau_{232}
\left[1-\exp\left(-{T_{\rm UP}\over\tau_{232}}\right)\right],
\ee
where $T_{\rm UP}$ is the period of uniform production prior to the 
formation of the solar system. The abundance ratio (Th/Eu)$_{{\rm ESS},r}$
is then
\be
\left({{\rm Th}\over {\rm Eu}}\right)_{{\rm ESS},r}=
\left({{\rm Th}\over {\rm Eu}}\right)_H
\left({\tau_{232}\over T_{\rm UP}}\right)
\left[1-\exp\left(-{T_{\rm UP}\over\tau_{232}}\right)\right],
\label{htheu}
\ee
where (Eu/H)$_{{\rm ESS},r}={\rm (Eu/H)}_Hf_HT_{\rm UP}$ 
has been used. With (Th/Eu)$_{{\rm ESS},r}=0.042/0.0917$ \cite{an89,ar99}, 
Eq. (\ref{htheu}) gives (Th/Eu)$_H=0.58$ for $T_{\rm UP}=10^{10}$ yr.
The value of (Th/Eu)$_H$ only decreases slightly to 0.54 if
$T_{\rm UP}=7\times 10^9$ yr is used instead.

For CS 22892--052, $\log({\rm Th/Eu})=-0.67\pm 0.11$ \cite{sn00}. 
Applying Eq. (\ref{ttheu}) with (Th/Eu)$_0={\rm (Th/Eu)}_H=0.58$ gives an 
age of $(20.1\pm 5.1)\times 10^9$ yr for this star (see also 
\cite{cow97,cow99}). However, the same procedure applied to CS 31082--001 
with $\log({\rm Th/Eu})=-0.22\pm 0.05$ \cite{hi02} leads to an unacceptable 
age of $(-0.9\pm 2.3)\times 10^9$ yr. Clearly, Eq. (\ref{ttheu}) must be 
applied to CS 22892--052 and CS 31082--001 with different values of 
(Th/Eu)$_0$. By using the age of $(13.5\pm 5.0)\times 10^9$ yr for
CS 31082--001 estimated above from the observed abundances of Th and U 
in this star, a value of (Th/Eu)$_0=1.20\pm 0.26$ is obtained from
Eq. (\ref{ttheu}). The value of (Th/Eu)$_H=0.58$ inferred above from 
Eq. (\ref{htheu}) should only represent some average over many $H$ events. 
The data for CS 31082--001 indicate that the relative yield of Th to Eu
in individual $H$ events can vary by a factor of $\sim 2$ around this 
average. Such variations
not only have severe impact on the Th/Eu chronology 
(e.g., \cite{gor99}), but also have profound implications for 
$r$-process nucleosynthesis (e.g., \cite{qi02}). Fission cycling
tends to result in a global steady flow between the fission
fragments and the fissioning nuclei, and hence, a robust yield pattern
of the heavy $r$-process elements Ba and above (e.g., \cite{frt99}).
This cannot occur during all the $H$ events in view of the variations
in the relative yield of Th to Eu. The possible role of fission (but
not fission cycling) in
producing a generic $H$ yield pattern and the associated variations
is discussed in \S\ref{fis}.

\subsection{Observational implications for the $r$-process\label{obsimp}}
The meteoritic data on $^{182}$Hf and $^{129}$I require diverse sources 
for the $r$-process and are consistent with the replenishment of these 
two nuclei in the ISM by the $H$ and $L$ events
at frequencies of $\sim (10^7\ {\rm yr})^{-1}$ and
$\sim (10^8\ {\rm yr})^{-1}$, respectively. These frequencies may be
explained by associating both kinds of events with core-collapse
supernovae (see \S\ref{met}). The solar $r$-process abundances
resulted from $\sim 10^3$ $H$ events and $\sim 10^2$ $L$ events. The
corresponding enrichment level by a single $H$ or $L$ event is in 
agreement with the observed $r$-process abundances in metal-poor stars
(see Fig. 9 and \cite{qi00}). The mass fraction of the heavy or light 
$r$-process elements in the solar system is
$X_{\odot,r}^{\rm heavy}\sim X_{\odot,r}^{\rm light}\sim 4\times 10^{-8}$
\cite{ar99}. If the amount of ISM that will mix with the ejecta from each 
event is $M_{\rm mix}\sim 3\times 10^4\,M_\odot$ as in the case of
supernovae,
$X_{\odot,r}^{\rm heavy}M_{\rm mix}\sim X_{\odot,r}^{\rm light}M_{\rm mix}
\sim 10^{-3}\,M_\odot$ of the heavy or light $r$-process elements must
be provided by the $\sim 10^3$ $H$ events or the $\sim 10^2$ $L$ events
that occurred prior to the formation of the solar system. This requires
$\sim 10^{-6}\,M_\odot$ or $\sim 10^{-5}\,M_\odot$ of the $r$-process
material be ejected by each $H$ or $L$ event, respectively. The
neutrino-driven wind from a protoneutron star can eject
$\sim 10^{-6}$--$10^{-5}\,M_\odot$ of material, although it remains to be
established that the conditions required for the $r$-process can be 
obtained in the wind (see \S\ref{snr}). The mass fraction of Fe in the
solar system is $X_\odot^{\rm Fe}\approx 10^{-3}$ \cite{an89}. 
In order to account for
$\sim 1/3$ of this Fe by the $\sim 10^2$ $L$ events that occurred prior to 
the formation of the solar system, each $L$ event must eject
$\sim (X_\odot^{\rm Fe}/300)M_{\rm mix}\sim 0.1\,M_\odot$ of Fe. This is in
agreement with the amount of Fe inferred from the light curves for a 
number of SNe II (see e.g., Table 1 of \cite{so02}). The rarity of
neutron star mergers appears to result in
conflicts with both the meteoritic data and observations of abundances
in metal-poor stars if they are associated with the $H$ or $L$ events. 
The arguments leading to this conclusion must be examined by more
sophisticated models of Galactic chemical evolution that include the
details for the mixing of the ejecta from individual nucleosynthetic
events with the ISM. While the possibility of neutron star mergers being 
an $r$-process source should also be studied, the discussion below again 
assumes that the 
$H$ and $L$ events are associated with core-collapse supernovae.

\subsubsection{three-component model for abundances in UMP and MP stars
and evolution of Eu abundance relative to Fe\label{eveufe}}
In the three-component model for abundances in UMP and MP stars
(see \S\ref{rhl}),
the enrichment level for a single $H$ or $L$ event corresponds to
$\log\epsilon_H({\rm Eu})\sim -2.5$ or [Fe/H]$_L\sim -2.5$. For a mixture
of a number $n_H$ of $H$ events and a number $n_L$ of $L$ events in the 
ISM, the corresponding abundances of Eu and Fe are given by
\bea
10^{\log\epsilon({\rm Eu})}&=&n_H\times 10^{\log\epsilon_H({\rm Eu})},
\label{neu}\\
10^{\rm [Fe/H]}&=&10^{{\rm [Fe/H]}_P}+n_L\times 10^{{\rm [Fe/H]}_L},
\label{nfe}
\eea
where [Fe/H]$_P\sim -3$ represents the $P$ inventory of Fe. On average,
a fraction $q=f_H/(f_H+f_L)$ of the $r$-process events is of
the $H$ kind ($q\sim 0.9$ for the three-component model). 
The probability for an ISM to have a mixture of
a number $n_H$ of $H$ events and a number $n_L$ of $L$ events is
\be
P(n_H,n_L)={(n_H+n_L)!\over n_H!n_L!}q^{n_H}(1-q)^{n_L+1}.
\label{pnhnl}
\ee
The 99\% confidence contours for $\log\epsilon({\rm Eu})$ over a range of 
[Fe/H] and the mean trend for evolution of Eu abundance relative to Fe 
can be calculated from Eqs. (\ref{neu})--(\ref{pnhnl}). The results for
$\log\epsilon_H({\rm Eu})=-2.48$, [Fe/H]$_P=-3$, [Fe/H]$_L=-2.48$, and
$q=10/11$ \cite{qi01} are compared with the data for metal-poor stars 
in Fig. 11. It can be seen that essentially all the data lie within the 
calculated 99\% confidence contours. The data are also in general agreement
with the calculated mean trend. There are several data below the calculated 
lower 99\% confidence contour at [Fe/H]~$> -1.5$ (see Fig. 11). 
This is because the calculations 
assumed that only the $L$ events added Fe beyond the $P$ inventory and
did not take into account the SN Ia contributions to Fe at 
[Fe/H]~$> -1.5$. Therefore, the observed $\log\epsilon({\rm Eu})$ values
over the wide range of $-3\lesssim [{\rm Fe/H}]\lesssim -1.5$ are in 
accord with the three-component model for abundances in UMP and MP 
stars. Note that essentially all the other models
(e.g., \cite{is99,fi02}) focused on explaining 
[Eu/Fe]~$=\log({\rm Eu/Fe})-\log({\rm Eu/Fe})_\odot$, but not
$\log\epsilon({\rm Eu})$, as a function of [Fe/H].

\begin{figure}[tb]
\begin{minipage}{2 cm}
\makebox[1cm]{}
\end{minipage}
\begin{minipage}{8 cm}
\epsfig{file=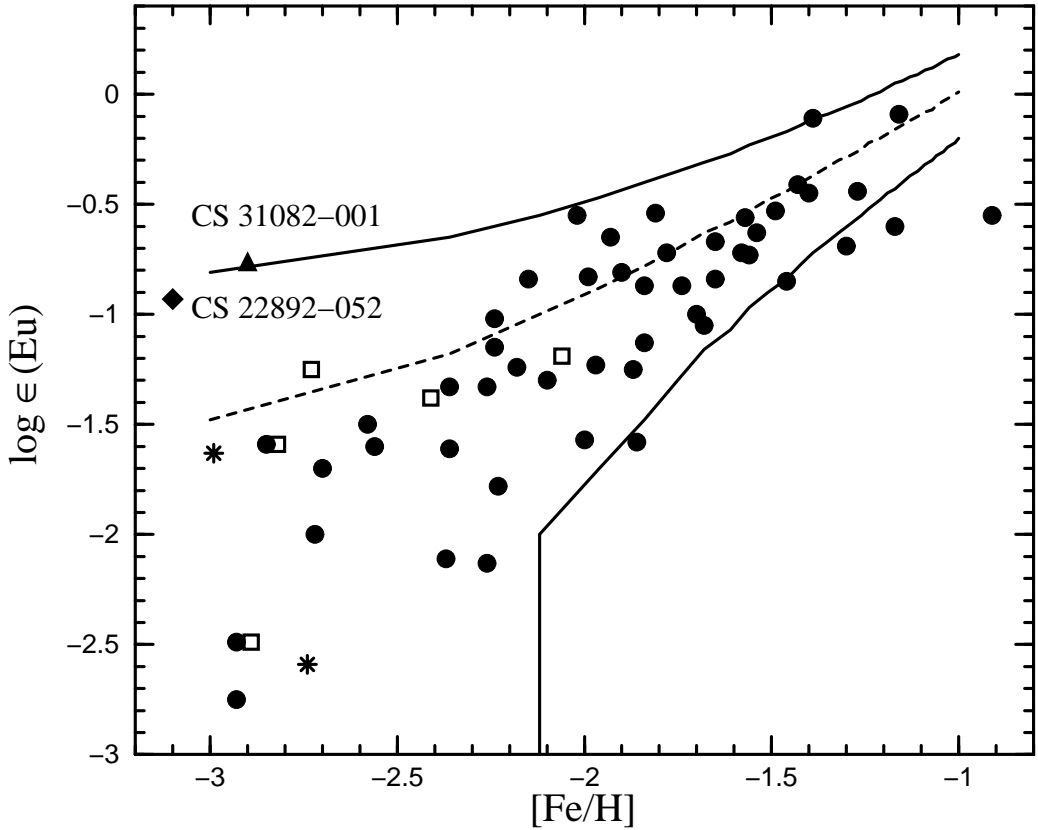,angle=270,scale=0.7}
\end{minipage}
\begin{center}
\begin{minipage}{16.5 cm}
\caption{Data on Eu at $-3\lesssim [{\rm Fe/H}]\lesssim -1$ (same symbols as
in Fig. 9) compared with the 99\% confidence contours (solid curves) and the 
mean trend (dashed curve) calculated from the three-component model 
\cite{qi01}. There
is general agreement between the model and the data. The several data below
the lower 99\% confidence contour at [Fe/H]~$> -1.5$ can be explained
by taking into account the Fe contributions from SNe Ia. The extremely high
Eu abundances in CS 31082--001 and CS 22892--052 are very close to the upper
99\% confidence contour and may not reflect the chemical composition of
the ISM from which these two stars formed. Instead, such high $r$-process
enrichments may be due to surface contamination by the ejecta from
core-collapse supernovae ($H$ events) in binaries \cite{qw01a}. See text for 
details.}
\end{minipage}
\end{center}
\end{figure}

The data for the UMP stars CS 31082--001
and CS 22892--052 lie very close to the calculated upper 99\% 
confidence contour for $\log\epsilon({\rm Eu})$ 
in the region where only the $H$ events could occur.
The observed value of $\log\epsilon({\rm Eu})=-0.76$ 
in CS 31082--001 \cite{hi02} or $\log\epsilon({\rm Eu})=-0.93$ 
in CS 22892--052 \cite{sn00} corresponds to $n_H=52$ or
35 $H$ events [$\log\epsilon_H({\rm Eu})=-2.48$]. 
With a frequency ratio of $f_H/f_L=10$, it is
unlikely that these many $H$ events could have occurred in an ISM without 
any accompanying $L$ events. Thus, the observed $r$-process abundances in
CS 31082--001 and CS 22892--052 may not reflect the composition of the ISM
from which they formed. Instead, each star may have had a binary companion, 
which evolved and exploded as a supernova $H$ event. The ejecta
from the explosion then contaminated the surface of CS 31082--001 or 
CS 22892--052 with high $r$-process enrichments \cite{qw01a}. This scenario
has two possible outcomes. If the binary survived the explosion, then the
highly $r$-process enriched star would have a neutron star or black hole
companion and exhibit periodic shifts in its radial velocity. Such shifts
were indicated by observations of CS 22892--052 \cite{pr01}. On the other
hand, if the binary was disrupted by the explosion, then the highly 
$r$-process enriched star might acquire a large space velocity from its
previous orbital motion. This may be revealed by proper motion measurements
of the star. 

\subsubsection{fission and the $H$ yield pattern\label{fis}}
It is important that observations be carried out to test the above 
surface contamination scenario
for explaining the highly $r$-process enriched metal-poor stars. This will
not only help establish core-collapse supernovae as the $r$-process site,
but also help determine the yield patterns of individual supernovae. If both
CS 31082--001 and CS 22892--052 are cases of surface contamination, then the
observed $r$-patterns in these two stars represent the yield patterns of two
individual $H$ events. Figure 8 shows that these two patterns are 
essentially the same in the region of the elements from Sr to Ir. 
However, the relative $H$ yields of Th to Eu inferred from the data for
CS 31082--001 and CS 22892--052 are very different (see \S\ref{chron}). 
Thus, the $H$ yield pattern has some generic feature but also exhibits 
certain variations. Concerning the generic $H$ yield pattern, there is a 
lack of data between Cd ($A=111$--114, 116) and Ba ($A=135$, 137, 138). 
However, the meteoritic data require that the $H$ events produce very 
little $^{129}$I (see \S\ref{met}). Observations also indicate that the Ba 
in the UMP star HD 140283 is 
dominantly $^{138}$Ba. This determination of the Ba isotopic composition
is possible because only the odd-$A$ isotopes $^{135}$Ba and $^{137}$Ba
have finite nuclear magnetic moments that give rise to hyperfine structure 
of the Ba spectra. The fraction of the odd-$A$ Ba isotopes in HD 140283 
was found to be $f_{\rm odd}=0.08\pm 0.06$ \cite{ma95} and $0.30\pm 0.21$ 
\cite{la02}. Thus, there appears to be a rather sharp rise in the $H$ 
yield at $^{138}$Ba. This is consistent with the very low $H$ yields at 
$A\sim 130$ required by the meteoritic data.

The above discussion of the $H$ yield pattern raises two issues: how can
the light ($A<130$) and heavy ($A>130$)
$r$-process elements be produced without any
significant production at $A\sim 130$ and how can some variations in the
yield pattern be accommodated? Both issues may be resolved by considering
fission of the progenitor nuclei during decay towards stability
after the $r$-process freezes out in core-collapse supernovae \cite{qi02}.
It is assumed that fission does not occur during the $r$-process
(i.e., no fission cycling) and that the $r$-process produces a
freeze-out pattern covering
$190\lesssim A<320$ with a peak at $A\sim 195$. As the progenitor nuclei
decay towards stability after the freeze-out, all of those with
$260\lesssim A<320$ eventually undergo spontaneous fission (e.g., 
\cite{ca01}). Due to the strong influence of the closed proton and neutron
shells at $^{132}$Sn, the fission of the progenitor nuclei with
$260\lesssim A<320$ is expected to
produce one fragment at $A\sim 132$ and the other at $130\lesssim A<190$.
Some of the progenitor nuclei with $230\lesssim A<260$ would also undergo
spontaneous fission during decay towards stability, thereby producing one
fragment again at $A\sim 132$ and the other at $100\lesssim A<130$. The
possibility of fission for the progenitor nuclei with 
$230\lesssim A<260$ may be greatly enhanced by
reactions with the neutrinos emitted in core-collapse supernovae as these
nuclei could be highly excited by such reactions (e.g., \cite{qi97}). 
Neutrino reactions may
even induce fission of the progenitor nuclei with $190\lesssim A<230$.
Experiments using energetic particles to induce fission of the stable or
long-lived nuclei in this mass range showed that the fission mode is
dominantly symmetric with no preference for a fission fragment at 
$A\sim 132$, and the mass ratio of the two fission fragments is 
$\sim 1$--1.2 (e.g., \cite{br63,mo01}). So neutrino-induced fission of
the progenitor nuclei with $190\lesssim A<230$ is expected to produce 
fragments at $86\lesssim A<125$.

In the above $r$-process scenario \cite{qi02}, 
nuclei with $86\lesssim A<190$ would be produced by 
fission of the progenitor nuclei with $190\lesssim A<320$ during decay 
towards stability after the freeze-out. The fission yields may be modified
by capture of the neutrons released from fission. For example, the 
accumulation of the fission fragments at $A\sim 132$ may be shifted
to somewhat higher $A$. This may explain the sharp rise in the $H$ yield
at $^{138}$Ba indicated by the isotopic composition of Ba in HD 140283.
In any case, there would be very little production
of the nuclei with $A\sim 130$. The ratio of the fission yields at
$86\lesssim A<190$ relative to the surviving abundances at 
$190\lesssim A<260$ depends on the number of neutrino reactions experienced
by each progenitor nucleus after the freeze-out. Variations of this number
among individual supernovae may explain the significant difference in
the abundance ratio Th/Eu between CS 31082--001 and CS 22892--052.
The progenitor nuclei far from stability deexcite mainly by neutron emission. 
The probability for deexcitation by fission increases as the charge of the 
progenitor nuclei increases during decay towards stability. Thus, 
neutrino-induced fission is closely associated with neutrino-induced
neutron emission. The level of neutrino interaction (a few neutrino
reactions per progenitor nucleus) after the freeze-out required to account
for the $H$ yield patterns is consistent with that for producing the nuclei
with $A=183$--187 by neutrino-induced neutron emission from the progenitor
nuclei in the yield peak at $A\sim 195$ (see Fig. 5) \cite{qi02}.

\subsubsection{supernova progenitors for $H$ and $L$ events\label{snp}}
The three-component model for abundances in UMP and MP stars 
(see \S\ref{rhl}) requires that the $H$ events produce very little of the
elements from Na to Ni including Fe \cite{qw02}. It is also argued above
that these events are associated with core-collapse supernovae, which
include SNe II from the collapse of O-Ne-Mg and Fe cores as well as
the silent supernovae from accretion-induced collapse (AIC) 
of white dwarfs. It is commonly thought that in SNe II, the elements below
Si are produced by hydrostatic burning during presupernova evolution and 
those from Si to Ni are mostly produced by explosive burning associated with
the shock propagation. In order not to produce
a significant amount of the elements from Na to Ni, an $H$ event
must either have a presupernova structure lacking substantial hydrostatic
burning shells or for some reason, have all the material below
the He burning shell fall back onto the central remnant \cite{qw02}. 
Stars with masses 
of $\sim 8$--$10\,M_\odot$ have very thin shells at the end of their lives 
and explode due to the collapse of an O-Ne-Mg core instead of an Fe core 
\cite{no84}. These stars could be the progenitors for the $H$ events. 
A supernova from O-Ne-Mg core collapse would only produce 
$\sim 0.002\,M_\odot$ of Fe \cite{ma88}. This is $\sim 50$ times smaller
than the Fe yield of $\sim 0.1\,M_\odot$ for an $L$ event.
In addition, the AIC of a bare white dwarf will not produce any of the 
elements from Na to Ni, and 
therefore, can also be associated with the $H$ events. 
A small amount of material would be ejected in
the neutrino-driven wind from the protoneutron star 
produced by any core collapse. This wind is a possible $r$-process 
site although it remains to be seen if adequate conditions for the 
$r$-process can be obtained in the wind (see \S\ref{snr}). 
Thus, both SNe II from O-Ne-Mg core collapse and AIC of white dwarfs may 
correspond to the $H$ events or at 
least a subset. Another possibility to accommodate the nucleosynthetic
requirements of the $H$ events is that SNe II from Fe core collapse of
progenitors at the higher-mass end
would suffer severe fall back \cite{wo95}.
It has been argued that SN 1997D, which ejected only $0.002\,M_\odot$ of 
Fe, was such a case \cite{tu98,be01}. 
A potential problem with such a scenario 
for the $H$ events is that the severe fall back may be an obstacle
for ejection of the $r$-process material from the inner most
regions of the supernova.

As mentioned earlier, the Fe yield of $\sim 0.1\,M_\odot$ assigned to the 
$L$ events is consistent with the amount of Fe inferred from the light
curves for a number of SNe II including SN 1987A 
(see e.g., Table 1 of \cite{so02}). In addition,
overabundances of Sr and Ba were observed in the spectra of SN 1987A
\cite{wi87}--\cite{maz95}. These were attributed to production by
the weak $s$-process during core He burning prior to the explosion
\cite{pr88}. However, the observed abundance ratio Ba/Sr is too large for
the weak $s$-process to explain. This favors an $r$-process origin for the
observed overabundances of Sr and Ba in SN 1987A \cite{ts01}. By assuming
a mass distribution identical to that of $^{56}$Ni inferred from the 
spectra and the light curve, it was estimated that SN 1987A produced
$\sim 6\times 10^{-6}\,M_\odot$ of Ba \cite{ts01}. For a dilution mass of
$\sim 3\times 10^4\,M_\odot$, this Ba yield corresponds to
$\log\epsilon({\rm Ba})\sim 0.2$ and is of the same order of magnitude
as the Ba yield assigned to an $L$ event 
[$\log\epsilon_L({\rm Ba})=-0.47$; see Table 1].
The observed abundance ratio Ba/Sr~$\sim 2.5({\rm Ba/Sr})_\odot$ in
SN 1987A \cite{maz92} is also consistent with the $L$ yield ratio
(Ba/Sr)$_L\sim({\rm Ba/Sr})_\odot$ (see Table 1 and \cite{an89}). 
The progenitor of this supernova was
a $20\,M_\odot$ star (e.g., \cite{wo87}). The progenitors of SN 1993J
\cite{sh94,ho96} and SN 1994I \cite{iw94} with approximately the same Fe 
yields as that of 
SN 1987A were $\sim 12$--$15\,M_\odot$ stars. The progenitor of SN 1997D
with an extremely low Fe yield of $0.002\,M_\odot$ might be a $26\,M_\odot$ 
star \cite{tu98}.
It is possible that the $L$ events are associated with SNe II from 
progenitors of $\sim 12$--$25\,M_\odot$. The SNe II from 
progenitors outside this mass range, as well as AIC of white dwarfs, 
may then be associated with the $H$ events.

\subsection{Other $r$-process studies based on abundances in metal-poor stars}
While only core-collapse supernovae and neutron star mergers are discussed
here as the possible $r$-process site, many alternatives were proposed in
the past (see \cite{co91} for a review). 
The various candidate sites differ in
the evolution timescale of the associated astrophysical object (e.g., SNe II
have short-lived progenitors) or in the dependence on the metallicity of the
associated astrophysical environment (e.g., some proposed sites rely on the
initial metal abundance for the seed nuclei or the neutron source). These
differences would
lead to different evolution of the $r$-process abundances in the
Galaxy. Thus, it is possible to discriminate the proposed $r$-process sites
based on observations of stellar abundances over a wide range of metallicity
\cite{ma92}. Most studies used data on [E/Fe] as a function of [Fe/H], where
[E/Fe]~$=\log({\rm E/Fe})-\log({\rm E/Fe})_\odot$ with E being an element
such as Eu. Interpreting the data at [Fe/H]~$< -2.5$ as indicating
an increase of [Eu/Fe] with increasing [Fe/H], some studies \cite{ma92,tr99}
claimed that Eu production was delayed compared with Fe production in the
early Galaxy. Assuming Fe production by most SNe II, these studies concluded
that the $r$-process occurs in SNe II from
longer-lived progenitors at the lower-mass end ($\sim 8$--$10\,M_\odot$).
With more data (see Fig. 9), it is clear that there is a large dispersion
in [Eu/Fe] rather than a definite trend at $-3\lesssim [{\rm Fe/H}]< -2.5$.
This dispersion is due to inhomogeneous chemical evolution
in the early Galaxy, 
where the abundances in an ISM were strongly influenced by
a small number of nucleosynthetic events. Thus, although SNe II from 
progenitors of $\sim 8$--$10\,M_\odot$ may be an $r$-process site, it is
inappropriate to justify this by assuming a smooth trend for evolution of
[Eu/Fe] at low [Fe/H].

The inhomogeneous nature of chemical evolution in the early Galaxy was
recognized by many authors (e.g., \cite{ry96,is99,fi02} and
\cite{au95}--\cite{arg02}).
Some authors \cite{ri99} followed Galactic dynamics and metal enrichment with
an $N$-body smooth particle hydrodynamic code, in which nucleosynthetic
products from a stellar particle were transferred to the neighboring gas
particles according to the distance between the stellar particle and the gas
particle. The observed large dispersion in [Ba/Fe] at [Fe/H]~$< -2.5$
similar to that in [Eu/Fe] was reproduced by this approach. The observed
large dispersions in [Sr/Fe], [Ba/Fe], and [Eu/Fe] at [Fe/H]~$< -2.5$
were also accounted for by a different approach that treated the chemical
evolution of the Galactic halo in terms of a series of cloud coalescence and
fragmentation \cite{tr01}. Yet another approach focused on the role of
individual supernovae. By assuming that stars form inside supernova remnants
with a chemical composition reflecting the mixture of the supernova ejecta
and the ISM swept by the remnant, it was shown that the large dispersion in
[Eu/Fe] at [Fe/H]~$< -2.5$ could be explained \cite{is99,fi02,arg00}.
Based on this approach, it was found that the data are consistent with
assigning Fe production to SNe II from progenitors of 
$>10\,M_\odot$ and Eu production to those from progenitors of either
8--$10\,M_\odot$ or $\geq 30\,M_\odot$ \cite{is99}. In any case,
not all SNe II can produce both Fe and Eu \cite{fi02}.

A more extreme approach assumed that the abundances in stars with
[Fe/H]~$< -2.5$ represent the yields of individual supernovae
\cite{sh98} (see also \cite{au95}). The progenitor of an SN II responsible
for the abundances in a star was identified by comparing the
observed [Mg/H] with the theoretical SN II yields of Mg, which are more or 
less consistently calculated by different models. The observed abundances
of the other elements in the star were then used to infer the corresponding
yields for the responsible SN II. The inferred Fe yields were in agreement
with the amount of Fe estimated from the light curves for a number of SNe II
\cite{sh98}. In addition, it was found that the SN II yields of Mn, Fe, Co,
and Ni inferred from the abundances in metal-poor stars increase while that
of Eu decreases with increasing progenitor mass \cite{ts98}. The observed
large dispersion in [Ba/Fe] at $-3\lesssim [{\rm Fe/H}]< -2.5$ was
attributed to a sharp decrease in the SN II yield of Ba from
$8.5\times 10^{-6}\,M_\odot$ for a $20\,M_\odot$ progenitor to
$4.5\times 10^{-8}\,M_\odot$ for a $25\,M_\odot$ progenitor \cite{ts00}.
Furthermore, the observed overabundances of Sr and Ba in SN 1987A 
(see \S\ref{snp}) were used to limit the progenitor masses
to $20\pm 0.7\,M_\odot$ for SNe II producing Ba and possibly 
$20\pm 3\,M_\odot$ for those producing Sr \cite{ts01}. While these results
are extremely quantitative and the underlying approach is rather interesting,
some problems with the adopted interpretation of the data for metal-poor
stars cannot be ignored. For example, the claimed sharp decrease in the 
SN II yield of Ba with increasing progenitor mass was based on interpreting
the data on [Ba/Fe] at $-3\lesssim [{\rm Fe/H}]< -2.5$ as indicating
a strict decrease of [Ba/Fe] with increasing [Fe/H]. However, in 
consideration of the observational uncertainties in [Fe/H], no clear trend
except for a large dispersion can be identified from e.g., the observed
values of [Ba/Fe]~$=0.9$, 0.18, and 1.17 for CS 22892--052 \cite{sn96}, 
HD 115444 \cite{we00}, and CS 31082--001 \cite{hi02} with [Fe/H]~$=-3.1$, 
$-2.99$, and $-2.9$, respectively.

All the above approaches to understand abundances in metal-poor stars
merit attention. In addressing the observed large dispersions in
$r$-process abundances at 
$-3\lesssim [{\rm Fe/H}]< -2.5$, most of these approaches assumed
that Fe is produced by the majority of SNe II while the $r$-process
elements are produced by a small subset. In contrast, the
three-component model discussed in \S\ref{rhl} attributed the Fe at
$-3\lesssim [{\rm Fe/H}]< -2.5$ to the $P$ inventory that was
dominantly produced by the very massive stars prior to the onset of major
formation of regular stars and assigned the subsequent Fe production by 
SNe II to the low-frequency $L$ events. The large dispersion in the
abundance of e.g., Eu at $-3\lesssim [{\rm Fe/H}]< -2.5$ was then 
caused by the occurrence of the high-frequency $H$ events producing Eu but 
no Fe \cite{wa00}. Some of the stars with [Fe/H]~$\sim -3$
such as CS 31082--001 and CS 22892--052 were argued to be cases of surface
contamination by the ejecta from the ($H$ event) explosion of a 
binary companion \cite{qw01a}. Future observations of more stars
with extreme $r$-process enrichments but low [Fe/H] along with the tests
of the surface contamination scenario discussed in \S\ref{eveufe}
should help identify the correct approach to
understand abundances in metal-poor stars.

\section{Summary and Outlook\label{end}}
There have been many advances in theoretical and observational studies of
the $r$-process since the last major review by Cowan et al.
\cite{co91} in 1991. This review tries to cover most of these advances.
A brief summary is given below with some comments on possible further 
progress.

The conditions required for the $r$-process have been examined in terms of 
the electron fraction, the entropy per baryon, and the dynamic timescale for
adiabatic expansion from high temperature and density (see \S\ref{cond}).
This parametrization of the astrophysical environment treats the 
determination of the neutron-to-seed ratio, the $r$-process, and the
freeze-out in a smooth sequence and is more realistic than the traditional
parametrization using a set of neutron number density, temperature, and
neutron irradiation time. The questions to be addressed are: how will
the conditions required for the $r$-process change if this process occurs
in an intense flux of supernova neutrinos (see \S\ref{nueff}) and is it
possible to refine these conditions by requiring the $r$-process to produce
the yield patterns inferred from meteoritic data and observations of 
metal-poor stars (see \S\ref{rhl})?

There have been many developments regarding core-collapse supernova and 
neutron star merger models of the $r$-process (see \S\ref{mod}). In 
particular, the neutrino-driven wind model associated with core-collapse
supernovae has been scrutinized in substantial detail. While current
models do not give the conditions for an $r$-process in the wind,
meteoritic data on $^{182}$Hf and $^{129}$I as well as observations 
of metal-poor stars are in support of
core-collapse supernovae being the $r$-process
site (see \S\ref{obs}). The wind model is attractive in that neutrino 
interaction can eject the right amount of material as required from 
each supernova to account for the solar $r$-process abundances.
The essential role of neutrinos in this model also provides a connection
between $r$-process nucleosynthesis and neutrino properties such as
flavor transformation. 
The crucial problem is: what remedies (see \S\ref{rem})
can lead to an $r$-process in the wind? 
As for the neutron star merger model, future
studies should try to reduce the uncertainties in the amount and the 
neutron-richness of the ejecta. The neutrino-driven wind associated 
with neutron star mergers (see \S\ref{simi}) is worth investigating
in detail. The observational consequences of the neutron star merger 
model, especially the challenges from meteoritic data and
observations of metal-poor stars (see \S\ref{obs}), need to be 
addressed. 

The evidence from meteoritic data and observations of metal-poor stars for
the diversity of $r$-process sources (see \S\ref{obs}) represents a 
breakthrough in $r$-process studies. The demonstration of some regularity 
in $r$-patterns and of large dispersions in $r$-process abundances at low 
[Fe/H] by stellar observations has had profound impact on the 
understanding of $r$-process nucleosynthesis (see \S\ref{obs}). Future
observations should study in more detail the regularities and the 
variations of $r$-patterns in metal-poor stars. In this regard, it is
extremely important to study the abundance ratios of the other heavy
$r$-process elements relative to Eu, such as Ba/Eu, La/Eu, and Nd/Eu,
at $-2.5\lesssim[{\rm Fe/H}]\lesssim -1.5$. If the heavy $r$-process
elements are indeed produced by a unique kind of events with a fixed
yield pattern, then these ratios should stay constant until the onset of
the $s$-process contributions at [Fe/H]~$\sim -1.5$. Extensive data
are currently available only for Ba, La, Nd, and Dy and exhibit
rather constant values of La/Eu and Dy/Eu, large spread in Nd/Eu with
no clear trend, and a systematic increase of Ba/Eu with increasing
[Fe/H] (see Fig. 4 of \cite{jo01} and Fig. 3 of \cite{qw01}). The
behavior of Ba/Eu and La/Eu at $-2.5\lesssim[{\rm Fe/H}]\lesssim -1.5$
indicates large contributions to Ba but no contributions to the next 
element La from another kind of $r$-process events associated with
Fe production (see \S\ref{rhl}). This needs to be tested by more 
extensive and precise data in the future. More data on Th and U
in metal-poor stars would be of great value to studies of variations
in the yield pattern of the heavy $r$-process elements 
(see \S\ref{chron}). A substantial data base for abundance patterns 
in very metal-poor stars with extreme $r$-process
enrichments may provide unique opportunities to study the $r$-process. 
Measurements of radial velocity and 
proper motion for such stars are crucial in testing the scenario of 
surface contamination by the ejecta from
core-collapse supernovae in binaries 
(see \S\ref{eveufe}). This is the
most promising approach to identify core-collapse supernovae as the
$r$-process site, although a more direct approach is to detect the $\gamma$
rays from the decay of $r$-process progenitor nuclei in supernova remnants
(e.g., \cite{qvw98,qvw99}). Another approach to demonstrate the
association of core-collapse supernovae with the $r$-process is to detect
overabundances of Sr and Ba in the spectra of individual events as in the 
case of SN 1987A (see \S\ref{snp}). The simultaneous observations of
light curves may also provide valuable information on the relationship
between Fe production and $r$-process nucleosynthesis in
core-collapse supernovae (see \S\ref{rhl}).

The $r$-process involves a strong interplay between nuclear physics and
astrophysics. For example, the conditions in the neutrino-driven wind 
associated with core-collapse supernovae are
determined by the characteristics of neutrino emission and the mass and
the radius of the protoneutron star, all of which are affected by the
properties of hot and dense matter. The amount and the neutron-richness
of the ejecta from neutron star mergers also depend on the behavior of
matter at high density. Thus, the equation of state at high density plays
a crucial role in both core-collapse supernova and neutron star merger
models of the $r$-process. This equation of state also affects the 
gravitational radiation from neutron star mergers \cite{oe02}, and 
therefore, may be constrained by the gravitational waves to be detected
by experiments such as LIGO. At present, almost all of the essential 
nuclear physical input for the $r$-process reaction network has to be
calculated from theory due to the lack of experimental data. This
situation may be changed by the Rare Isotope Accelerator (RIA), which
is planned as the next major facility for nuclear physics in the
United States \cite{nsac02}. Similar experimental efforts at RIKEN in
Japan and GSI in Germany are also being discussed. Some of the issues 
regarding the $r$-process that may be addressed by RIA and similar
facilities are considered below.

The assumption of $(n,\gamma)\rightleftharpoons(\gamma,n)$ equilibrium 
greatly simplifies the reaction network for the $r$-process and provides
a useful framework to discuss this process. It would be of great help 
to $r$-process calculations for different astrophysical environments
if the conditions for $(n,\gamma)\rightleftharpoons(\gamma,n)$ 
equilibrium (see \S\ref{nggn}) can be 
established by experimental data. The masses, the $\beta$-decay 
lifetimes, and the neutron-capture cross sections for neutron-rich nuclei 
far from stability are needed for this purpose and may be obtained from 
experiments at RIA and similar facilities.

Calculations with constant \cite{kr93} or time-dependent \cite{fr99} 
neutron number density and temperature showed that an overstrong
$N=82$ shell causes severe underproduction of the nuclei with 
$A\sim 110$--126. Quenching the strength of the $N=82$ shell greatly
alleviates the problem \cite{ch95}. As the $N=82$ shell must be 
sufficiently strong to produce the $r$-process abundance peak at 
$A=130$, quenching should be most effective far below $A=130$. Thus,
the deficiency at the upper end of $A\sim 110$--126 may require
other remedies such as neutrino-induced neutron emission from the
progenitor nuclei in the abundance peak at $A=130$ (see \S\ref{nueff}).
The exact form of quenching may be identified by measurements of masses 
for the nuclei with $A\sim 110$--126 and $N\leq 82$ at RIA
and similar facilities.

Meteoritic data and the observed $r$-patterns in the ultra-metal-poor
stars CS 22892--052 and CS 31082--001 suggest that the high-frequency
$r$-process events should produce the nuclei on both sides of 
$A\sim 130$ but very little of those with $A\sim 130$ (see \S\ref{obs}).
As a plausible scenario to explain this peculiar yield pattern, it 
was assumed that the $r$-process produces a freeze-out pattern 
covering $190\lesssim A<320$ 
with a peak at $A\sim 195$. Neutrino-induced fission of the progenitor
nuclei during decay towards stability then produces the nuclei with
$86\lesssim A<190$ but very little of those with $A\sim 130$ due to
the characteristics of fission yields (see \S\ref{fis}). The competition 
between neutron emission and fission in the deexcitation of unstable 
neutron-rich nuclei and the fission yields of such nuclei are crucial 
to the above scenario. Experiments at RIA and similar facilities
may shed some light in this regard. 

A full understanding of the $r$-process requires quantitative 
description of the astrophysical environment, reliable nuclear
physical input, and correct interpretation of meteoritic data
and stellar observations. If the role of neutrinos in the $r$-process
can be established, then neutrino properties are an essential 
part of the problem. In view of the planning for measurements of nuclear
properties at RIA and similar facilities, the ongoing 
observations of metal-poor stars with large telescopes
(e.g., the Hubble Space Telescope, the Keck Telescopes, and the
Very Large Telescope), and the ongoing and planned experiments on 
neutrino oscillations (e.g., Super Kamiokande, Sudbury Neutrino
Observatory, KamLAND, Mini-BooNE, and MINOS), much further progress 
in the understanding of the $r$-process is expected in the future.

\section*{Acknowledgments}
I would like to thank Al Cameron, George Fuller, Wick Haxton, 
Karlheinz Langanke, Friedel Thielemann, Petr Vogel, Jerry Wasserburg, 
and Stan Woosley for enriching my education regarding the $r$-process.
Comments by the two referees (one being Friedel Thielemann), Jerry 
Wasserburg, and Stan Woosley have greatly improved this review.
I deeply appreciate the indulgence from the editor, Professor Amand 
Faessler. This work was supported in part by the US Department of 
Energy under grants DE-FG02-87ER40328 and DE-FG02-00ER41149.

\end{document}